\documentclass[journal,12pt,onecolumn,draftclsnofoot]{IEEEtran}
\IEEEoverridecommandlockouts

\usepackage{cite}
\usepackage{amsmath,amssymb,amsfonts}
\usepackage{algorithm}
\usepackage{algorithmic}
\usepackage{graphicx,psfrag}
\usepackage{graphicx}
\usepackage{textcomp}
\usepackage{xcolor}
\usepackage{amsthm}
\usepackage[justification=centering]{caption}
\usepackage{overpic}
\usepackage{multirow}
\usepackage{fancyhdr}
\usepackage{lmodern}

\title{Space-time Coded Differential Modulation for Reconfigurable Intelligent Surfaces}
\author{Jiawei Qiu \footnote{Jiawei Qiu was with the Department of Electrical and Computer Eng. McGill University, Montreal QC Canada. He is now with INRS Telecomunications, University of Quebec, Montreal QC Canada.}
 and
 Harry Leib \footnote{Harry Leib is with the Department of Electrical and Computer Eng., McGill University, Montreal, QC Canada, Email: harry.leib@mcgill.ca}	}

\begin{document}
	\maketitle	

\vspace*{-20mm}

	\begin{abstract}
		Reconfigurable Intelligent Surfaces (RIS)  hold the  promise of improving significantly coverage, as well as  spectral and energy efficiency in wireless communication systems. Techniques  based on RIS form a key technology for 6G systems. An important  issue in RIS technology is Channel State Information (CSI), which is much more difficult to acquire in such systems.. This work introduces a Differential Space-Time Modulation (DSTM) scheme integrated with  Differential Reflecting Modulation (DRM) to bypass the requirement for CSI in such systems, while providing error rate gains. The DSTM scheme is based on unitary group codes.  We first consider  uncoded DRM for RIS to serve as a reference point. Next we provide an overview of  DSTM and outline the procedures for its integration with DRM. Furthermore, we explore the extension of both the original DRM and the coded DRM-DSTM scheme  to a larger number of  RIS reflecting patterns $K$, and  provide tables of codes for $K= 2, 3, 4$. Encoding and decoding complexities are studied as well. Extensives simulation results over quasi-static Rayleigh fading channels   confirm the effectiveness of the DRM-DSTM coded system, illustrating its advantages over uncoded DRM with proper system parameters.
		
	\end{abstract}


	\section{Introduction}
	Reconfigurable Intelligent Surface (RIS) technology constitues a powerful approach for creating favorable wireless communication environments~\cite{RISee,MISOIRS,smart,10418182,  an2024}. As a result, the  RIS technology emerges as  a key enabler for  6G telecommunications \cite{wu2024intelligent, sode2024reconf, shi2024ris, wang2023road, 10904090}   An RIS consists of a large number of passive reflecting elements, each capable of altering incident signals by adjusting their phase or amplitude ~\cite{zhao2019survey}. These elements are made from electromagnetic materials, with size and inter-distance much smaller than the signal wavelength. Unlike traditional relays, RIS can be implemented with minimal hardware complexity and cost, relying on low-power and low-complexity electronic circuits~\cite{RISvsRelay}. A key feature of RIS is its reconfigurability, allowing it to modify the wireless environment after deployment, thus making  the electromagnetic propagation environment programmable and controllable~\cite{SmartRIS}.

	Recently, there has been growing interest in RIS-based communication systems. Significant research has been conducted on RIS-based  modulation techniques~\cite{basar2019,basar2020,basar2019wireless,RM}, as well as on CSI acquisition~\cite{alwazani2020,he2019,mishra2019}. In particular, Basar in~\cite{basar2020} expanded the use of RIS to index modulation (IM) by introducing RIS-space shift keying (RIS-SSK) and RIS-spatial modulation (RIS-SM) schemes. These techniques can improve spectral efficiency by intelligently reflecting signals and optimizing the selection of receive antenna indices. To address the challenge that, large intelligent metasurfaces (LIM), equipped with a large number of low-cost metamaterial antennas, can only passively reflect signals through specific phase shifts, He \textit{et al} in~\cite{he2019} proposed a comprehensive method for estimating the cascaded channels between the transmitter-LIM and LIM-receiver. However, most of the proposed RIS-based schemes still assume perfect CSI for detection~\cite{perfectCSI1,perfectCSI3,perfectCSI4,DRM}. Channel estimation in RIS-assisted systems, which involves the transmitter-RIS, RIS-receiver, and direct transmitter-receiver channels, remains a complex task due to the lack of baseband signal processing at the RIS units. Despite efforts such as in ~\cite{imperfectCSI1,imperfectCSI2,alwazani2020} to solve this issue, CSI acquisition remains a significant challenge in RIS systems.

	Noncoherent detection, a demodulation technique that does not require CSI, reduces system complexity and offers greater reliability than coherent detection where CSI is necessary~\cite{noncobetter}. As a result, noncoherent detection has gained attention in RIS-assisted systems~\cite{nonco1,nonco2,nonco3,nonco4,nonco5}. Kun \textit{et al} in~\cite{nonco2} proposed a noncoherent demodulation approach using differential decoding with random phase configurations at the RIS for RIS-assisted Orthogonal Frequency Division Multiplexing (OFDM) systems. This technique eliminates the need for channel estimation and is surpassing coherent detection in error performance. Ma \textit{et al} in~\cite{nonco4} introduced two RIS-assisted $M$-ary frequency-modulated differential chaos shift keying schemes, offering low-cost, low-power, and high-reliability solutions. These schemes not only reduce the need for CSI but also demonstrate improved performance in simulations compared to other systems.

	Over the past two decades, space-time coding and modulation have significantly enhanced spectral efficiency and power consumption  by using multiple antenna technologies~\cite{Foschini1996,Onggosanusi2002,Verde2013}. Various types of space-time codes were introduced in~\cite{tarokh1998}, where trellis codes were applied to improve communication reliability and data rates in fading channels with multiple transmit antennas. These codes provide an effective balance between data rate, diversity gain, and trellis complexity. Space-time block codes were also presented in works such as~\cite{Alamouti1998,Tarokh1999}. Alamouti, in~\cite{Alamouti1998}, proposed a two-branch transmit diversity scheme, which achieves the same diversity order as maximal-ratio receiver combining (MRRC) using one transmit and two receive antennas. 
	
	Similar to RIS-based modulation, much of the research on space-time coding has been based on the assumption that perfect CSI are available at the receiver. However, in some cases, CSI is preferred to be avoided to reduce cost and complexity. In scenarios where fading conditions change rapidly, CSI becomes significantly difficult to obtain or may require an excessive number of training symbols. To address such challenges, differential modulation techniques were developed, allowing for demodulation with CSI being unknown, similar to differential phase-shift keying (DPSK) for a single transmitter. Hughes, in~\cite{DSTM}, proposed a space-time differential modulation technique based on group codes that can be applied to any number of transmitters and receivers, enabling demodulation with or without channel estimates. Hochwald and Sweldens, in~\cite{DUSTM} introduced the unitary space-time block codes. These codes are ideal for Rayleigh fading channels where the fading coefficients are unknown to neither the transmitter nor the receiver, and are also suitable for piecewise-constant fading models~\cite{Systematic}. Zhao \textit{et al} in~\cite{Zhao2008}, presented a differential space-time modulation system with adaptive eigen-beamforming and power loading, based on the correlation of transmitter-side channels. Their analysis and simulations showed that the proposed scheme, designed for spatially correlated Rayleigh fading channels, significantly outperformed conventiona Differential Space-Time Modulation (DSTM) systems by taking changes in channel conditions and total transmit power into account.
	
	Acknowledging the benefits of eliminating the need for channel estimates in both RIS and space-time coding systems, Differential Reflecting Modulation (DRM)~\cite{DRM} and DSTM)~\cite{DSTM} were developed, both using a small number of symbol time slots and antennas. Different from~\cite{nonco1,nonco2,nonco3,nonco4,nonco5}, the DRM scheme primarily takes advantage of $M$-ary PSK, mapping information bits to the activation order of reflecting patterns and the phases of transmitted signals. This method enables differential detection and reduces the need for channel estimation. The unitary group codes-based DSTM, introduced in~\cite{DSTM}, is designed for systems with two transmit antennas and can be demodulated without CSI. To the best of our knowledge, we have not found the integration of DSTM into DRM in the literature.

	In this work we integrate  space-time coded differential modulation into DRM  to bypass the requirements of CSI. The fundamentals of both schemes are demonstrated , including the encoding process, signal transmission, detection method, and reflecting pattern selection. By using extensive computer simulation results we demonstrate  the superiority of the coded scheme. Then, we extend both schemes to higher sizes by deriving general expressions that can be applied to any number of transmitting and receiving antennas. With such an extension, we construct a general coded system, including cyclic and dicyclic group codes integrated into DRM. Finally, the error performance and time complexity are compared for the two proposed schemes. 
	
	The rest of the paper is organized as follows. In Section \ref{section2}, we introduce the basic model of the uncoded DRM  and the DSTM schemes, and the combination of these over only two symbol time slots, along with a performance comparison. Section \ref{section3} depicts  expressions of the space-time codes with a group structure, and provides tables of cyclic and dicyclic group codes used in the encoding processes, resulting in  a general coded system. Section \ref{section4} presents the bit error rate (BER) results versus the signal-to-noise ratio (SNR) for the coded and uncoded systems. Conclusions are drawn in the last section.
	
	\section{System Model of DRM and DSTM Basics}\label{section2}

	\begin{figure}[!th]
		\centering
\includegraphics[width=0.6\textwidth]{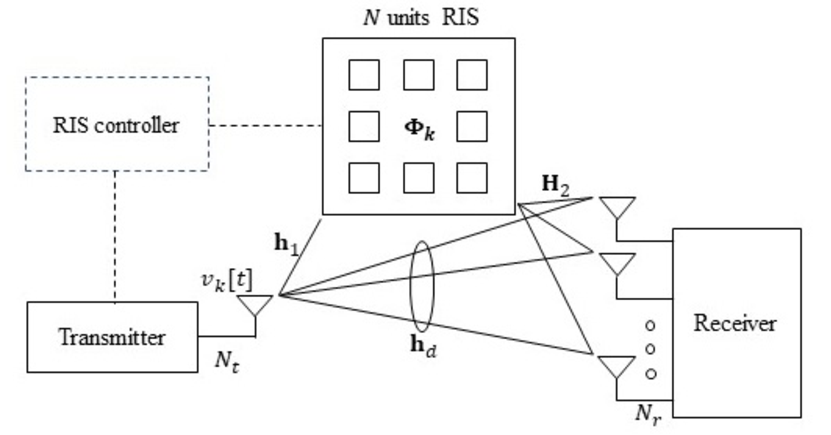}
		\caption{A RIS-based communication system.}
		\label{RIS system}
	\end{figure}
	
	In this section, we start by  presenting the system model of DRM from \cite{DRM} for reader convenience, in order to establish a reference point to our work that cosists of DRM with DSTM coding. We continue with the basic properties of DSTM, which is then combined with DRM forming DSTM-DRM.  A comparison of the error performance of DSTM-DRM and DRM is also presented.
	
	\subsection{System Model for DRM}\label{section2.1}
	This work considers a communication system based on RIS technology as in \cite{DRM} consisting of  $N_t$ transmit and $N_r$ receive antennas, where only one transmit antenna is active at a time, as illustrated in Fig. \ref{RIS system}. The RIS is equipped with $N$ reflecting units forming reflecting patterns. A reflecting pattern is characterized by a diagonal matrix $\mathbf{\Phi}_k \in \mathbb{C}^{N\times N}$, which controls the activation and phase shifts of all the reflecting units. It is assumed that there are $K$ reflecting pattern candidates for transmission, represented by the set $\Psi=\{\mathbf{\Phi}_1,\mathbf{\Phi}_2,\cdots,\mathbf{\Phi}_K\}$. In practice, RIS elements are configured to optimize signal reflection. As a result, the diagonal entry $(\mathbf{\Phi}_k)_{n,n}$ of a reflecting pattern $\mathbf{\Phi}_k \in \Psi$ has an amplitude of either 1 or 0, depending on whether the $n$-th reflecting unit is active. Mathematically, the amplitude is defined as $(\beta_k)_n\in\{0,1\}$, and when $\mathbf{\Phi}_k$ is active (i.e when $(\beta_k)_n =1$)  $(\mathbf{\Phi}_k)_{n,n} \in \{0\} \cup \{{\rm exp}(j(\theta_k)_n)\}, 1\leq n\leq N$, where $(\theta_k)_n \in [0,2\pi]$ denotes the phase shift angle. Additionally, in Fig. \ref{RIS system}, $\mathbf{h}_1 \in \mathbb{C}^{N\times1}$ represents the channel vector between the transmitter and RIS, $\mathbf{H}_2 \in \mathbb{C}^{N_r\times N}$ is the channel matrix between the RIS and the receiver, and $\mathbf{h}_d \in \mathbb{C}^{N_r\times1}$ is the channel vector for the direct link. These channels are modeled as uncorrelated quasi-static Rayleigh fading, each represented by a random variables $h$, which is zero-mean complex Gaussian with  normalized power, $E\{|h^2\}=1$~\cite{DRM,uncorrelated1,uncorrelated2}.

	\begin{figure}[htbp!]
		\centering
		\includegraphics[width=0.8\textwidth]{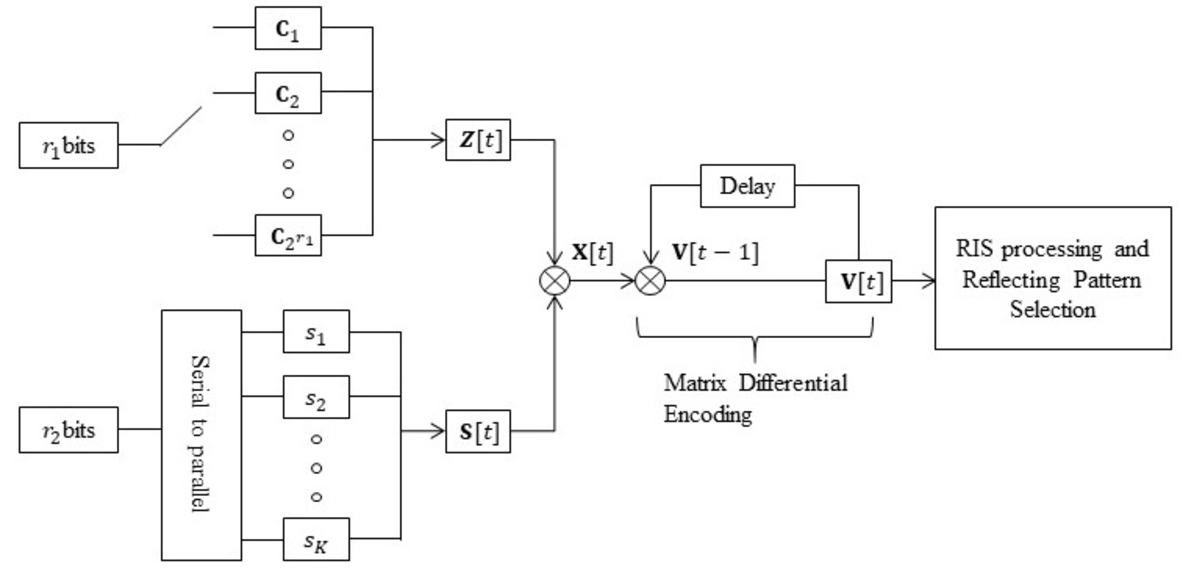}
		\caption{DRM encoding.}
		\label{DRMencoding}
	\end{figure}
	
	\subsubsection{Encoding Process}\label{section:Encoding}
\begin{figure}[htbp!]
		\centering
		\includegraphics[width=0.5\textwidth]{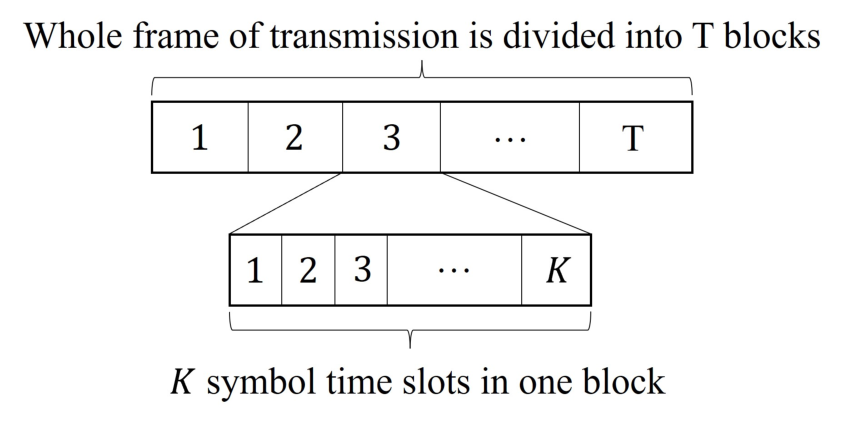}
		\caption{The structure of a transmission frame.}
		\label{block}
	\end{figure}

	While in \cite{DRM} it is mentioned that encoding can be done jointly at the transmitter and RIS or separately, in our work we assume that the encoding
	process is done at the tramsmitter, while RIS specific operations (such as reflecting patter selection) are done at the RIS using information from the transmitter which requires an RIS controller aiding this function. The encoding process is depicted in Fig. \ref{DRMencoding}. The entire frame is divided into $T$ blocks, with each block containing $K$ symbol time slots, as shown in Fig. \ref{block}. With such structure during the $t$-th block, $K$ reflecting pattern candidates are activated in a specific sequence, occupying the $K$ symbol time slots, along with the $M$-PSK symbols~\cite{DRM}. For each block, the number of information bits is given by $r=r_1+r_2=\lfloor {\rm log}_2K!\rfloor+K{\rm log}_2M$, where both $r_1$ and $r_2$ are integers. The first $r_1=\lfloor {\rm log}_2K!\rfloor$ bits are mapped to a $K\times K$ permutation matrix $\mathbf{Z}[t]$, which determines the activation sequence of the reflecting patterns. During the block, the activation sequence of the $K$ reflecting pattern candidates is selected from all possible permutations. While $K!$ permutations are theoretically possible, not all of them are required. For example, in a two-bit encoding scheme with $K=3$ symbols per block, as shown in Table \ref{tb:permu}, only four matrices are needed to represent the two-bit binary codes, instead of all six permutations. In this case, with two information bits there is need for four permutation matrices for $K=3$ time slots. When two time slots are used, there are only two possible permutations, each mapped to one binary bit, either $0$ or $1$. Thus, $\lfloor {\rm log}_2K!\rfloor$ is an upper bound, and the number of permutation matrices is $2^{r_1}$. These matrices are denoted as $\mathbf{C}_1,\mathbf{C}_2,\cdots,\mathbf{C}_{2^{\lfloor {\rm log}_2K!\rfloor}}$, with one being mapped to $\mathbf{Z}[t]$ based on the first $r_1$ information bits.
	
	\begin{table}[!h]
		\centering
		\begin{tabular}{c|c}
			\hline
			$r_1=2$ Bits & $3\times 3$ permutation matrix $\mathbf{Z}[t]$\\
			\hline
			$00$ & $\mathbf{C}_1=\begin{bmatrix}
				0 & 1 & 0\\[-2pt]
				0 & 0 & 1\\[-2pt]
				1 & 0 & 0
			\end{bmatrix}$\\
			\hline
			$01$ & $\mathbf{C}_2=\begin{bmatrix}
				0 & 1 & 0\\[-2pt]
				1 & 0 & 0\\[-2pt]
				0 & 0 & 1
			\end{bmatrix}$\\
			\hline
			$10$ & $\mathbf{C}_3=\begin{bmatrix}
				1 & 0 & 0\\[-2pt]
				0 & 1 & 0\\[-2pt]
				0 & 0 & 1
			\end{bmatrix}$\\
			\hline
			$11$ & $\mathbf{C}_4=\begin{bmatrix}
				
				1 & 0 & 0\\[-2pt]
				0 & 0 & 1\\[-2pt]
				0 & 1 & 0
			\end{bmatrix}$\\
			\hline
		\end{tabular}
		\caption{Bit encoding with $r_1=2$ and $K=3$}
		\label{tb:permu}
	\end{table}

	The remaining $r_2 = K\log_2M$ bits are mapped to $K$ $M$-PSK symbols, denoted as $s_k$ for $1 \leq k \leq K$. These $K$ symbols are arranged into a $K \times K$ diagonal matrix, $\mathbf{S}[t] = {\rm diag}(s_1, \cdots, s_K)$. An information-conveying matrix $\mathbf{X}[t] \in \mathbb{C}^{K \times K}$ is then formed as
	\begin{align}
		\mathbf{X}[t]=\mathbf{Z}[t]\mathbf{S}[t]. \label{eq:InforCarry} 
	\end{align}
	From $\mathbf{X}[t]$, a matrix $\mathbf{V}[t]$ can be generated after matrix differential encoding as
	\begin{align}
		\mathbf{V}[t]=\mathbf{V}[t-1]\mathbf{X}[t], \label{eq:diffencode}
	\end{align}
	where $\mathbf{V}[t-1]$ is generated in the previous block. The first block in a frame contains  no information and hence $\mathbf{X}[0]=\mathbf{I}$. Appendix \ref{apdx1} shows that $\mathbf{V}[t]$ is the multiplication of a permutation matrix and a diagonal matrix. Hence  $\mathbf{V}[t]=\tilde{\mathbf{Z}}[t]\tilde{\mathbf{S}}[t]$, where $\tilde{\mathbf{Z}}[t]$ is a permutation matrix and $\tilde{\mathbf{S}}[t]$ denotes a diagonal matrix.
	We can further express (\ref{eq:diffencode}) as
	\begin{align}
		\mathbf{V}[t]=\tilde{\mathbf{Z}}[t]\tilde{\mathbf{S}}[t]=\left[ \begin{array}{ccc}
			\tilde{\mathbf {z}}_1[t] & \cdots & \tilde{\mathbf {z}}_K[t]
		\end{array}
		\right]\tilde{\mathbf{S}}[t],\label{eq:tildeZS}
	\end{align}
	where $\tilde{\mathbf{z}}_k[t]\in\mathbb{C}^{K\times 1},1\leq k\leq K$ represents the $k$-th column of $\tilde{\mathbf{Z}}[t]$.
	
	\subsubsection{Signal Model}\label{section:SM}
	Since the column vector $\tilde{\mathbf{z}}_k[t]$ of the permutation matrix $\tilde{\mathbf{Z}}[t]$ contains exactly one nonzero entry, which is 1, it can be viewed as an $i$-th standard  basis vector. Specifically, if the $i$-th element of $\tilde{\mathbf{z}}_k[t]$ is 1, then $\mathbf{e}_i[t]=\tilde{\mathbf{z}}_k[t]$. When a permutation matrix is multiplied by a diagonal matrix $\tilde{\mathbf{S}}[t]$, the $k$-th diagonal element $\tilde{s}_k[t]$ is shifted to the $i$-th row of $\tilde{\mathbf{z}}_k[t]$, which implies
	\begin{align}
		\mathbf{V}[t] &= {\left[ \begin{array}{ccc}
				\tilde{\mathbf {z}}_1[t] & \cdots & \tilde{\mathbf {z}}_K[t]
			\end{array}
			\right]} {\left[ \begin{array}{ccc}
				{\tilde{s}_{1}[t]} & \cdots & 0 \\
				\vdots & \ddots & \vdots \\
				0 & \cdots & {\tilde{s}_{K}[t]}
			\end{array}
			\right]}\\
		& = {\left[ \begin{array}{ccc}
				\tilde{\mathbf {z}}_1[t]{\tilde{s}_{1}[t]} & \cdots & \tilde{\mathbf {z}}_K[t]{\tilde{s}_{K}[t]}
			\end{array}
			\right]}={\left[ \begin{array}{ccc}
				\mathbf {v}_{1}[t] & \cdots & \mathbf {v}_{K}[t]
			\end{array}
			\right ]}, \label{eq:Vt}
	\end{align}
	where $\mathbf{v}_k[t]=\tilde{\mathbf{z}}_k[t]\tilde{s}_k[t]=\mathbf{e}_i[t]v_k[t]$. The $k$-th column vector $\mathbf{v}_k[t]$ contains the information for the transmission during the $k$-th time slot of the $t$-th block. In this context, $\mathbf{e}_i[t]$ is a vector of size $K \times 1$ where only the $i$-th element is 1, indicating that the RIS activates the $i$-th reflection pattern $\mathbf{\Phi}_i$ because $\mathbf{e}_i[t]=\tilde{\mathbf{z}}_k[t]$. The nonzero element $v_k[t]$ in the $i$-th row of $\mathbf{v}_k[t]$ is an $M$-PSK symbol $\tilde{s}_k[t]$.
	
	From Fig. \ref{RIS system}, the symbol $v_k[t]$ reaches the receiver in two ways during the $k$-th time slot of the $t$-th block. The first path is through the direct link, where $\mathbf{h}_d\in\mathbb{C}^{N_r\times 1}$ acts on the transmitted symbol. The second path involves the signal passing through the transmitter-RIS channels, getting reflected by the RIS, and then traveling through the RIS-receiver channels. This process is represented by the product $\mathbf{H}_2\mathbf{\Phi}_i\mathbf{h}_1\in\mathbb{C}^{N_r\times 1}$ which operates on the transmitted symbol.
	As a result, the signal received during the $k$-th time slot of the $t$-th block in the frame is \cite{RM, DRM}
	\begin{align}
		\mathbf{y}_k[t]=(\mathbf{h}_d+\mathbf{H}_2\mathbf{\Phi}_i\mathbf{h}_1)v_k[t]+\mathbf{n}_k[t], \label{eq:recivedvector}
	\end{align}
	where $\mathbf{n}_k[t]$ denotes  uncorrelated complex Gaussian noise vector with zero mean and variance $\sigma^2\mathbf{I}_{N_r}$.
	
	For convenience we stack all the received and transmitted symbols  in one block, and introduce the following matrices,
	\begin{align} 
		\tilde {\mathbf {H}}_{d}&=[\overbrace {\mathbf {h}_{d},\mathbf {h}_{d},\cdots, \mathbf {h}_{d}}^{K}]\in \mathbb {C}^{N_{r}\times K}, \label{eq:Hd} \\
		\tilde {\mathbf {H}}_{2}&=[\overbrace {\mathbf {H}_{2},\mathbf {H}_{2},\cdots, \mathbf {H}_{2}}^{K}]\in \mathbb {C}^{N_{r}\times KN}, \label{eq:H2} \\
		\tilde {\mathbf {H}}_{1}&=\overbrace { \begin{bmatrix} ~\mathbf {h}_{1} &\quad \mathbf {0} &\quad \cdots &\quad \mathbf {0}\\ ~\mathbf {0} &~\mathbf {h}_{1}& \cdots &\quad \mathbf {0}~\\ ~\vdots &\quad \vdots &\quad \ddots &\quad \vdots ~\\ ~\mathbf {0} &\quad \mathbf {0} &\quad \cdots &\quad \mathbf {h}_{1}\\ \end{bmatrix}}^{K}\in \mathbb {C}^{ KN\times K},\label{eq:H1} \\
		\mathbf{Q}&=\begin{bmatrix} ~\boldsymbol{\Phi }_{1} &\quad \mathbf {0} &\quad \cdots &\quad \mathbf {0}\\ ~\mathbf {0} &\quad ~\boldsymbol{\Phi }_{2}&\quad \cdots &\quad \mathbf {0}~\\ ~\vdots &\quad \vdots &\quad \ddots &\quad \vdots ~\\ ~\mathbf {0} &\quad \mathbf {0} &\quad \cdots &\quad \boldsymbol{\Phi }_{K}\\ \end{bmatrix}\in \mathbb {C}^{KN\times KN}. \label{eq:Q}
	\end{align}
	The equivalent channel matrix $\mathbf{H}\in\mathbb{C}^{N_r\times K}$ becomes
	\begin{align}
		\mathbf{H}&=\tilde {\mathbf {H}}_{d}+\tilde {\mathbf {H}}_{2}\mathbf {Q}\tilde {\mathbf {H}}_{1} \label{eq:Htotal}\\
		&=
		\begin{bmatrix}
			\mathbf{h}_d+\mathbf{H}_2\mathbf{\Phi}_1\mathbf{h}_1 & \mathbf{h}_d+\mathbf{H}_2\mathbf{\Phi}_2\mathbf{h}_1 & \cdots & \mathbf{h}_d+\mathbf{H}_2\mathbf{\Phi}_K\mathbf{h}_1
		\end{bmatrix}. \label{eq:H}
	\end{align}
	Therefore,  $\mathbf{y}_k[t]$ from (\ref{eq:recivedvector}) corresponds to the $k$-th column of the received signal matrix $\mathbf{Y}[t]\in\mathbb{C}^{N_r\times K}$ in block $t$, where $\mathbf{Y}[t]=[\mathbf{y}_1[t], \mathbf{y}_2[t], \dots, \mathbf{y}_K[t]]$, which can be expressed as
	\begin{align}
		\mathbf {Y}[t]=&\begin{bmatrix}
			(\tilde {\mathbf {H}}_{d}+\tilde {\mathbf {H}}_{2}\mathbf {Q}\tilde {\mathbf {H}}_{1})\mathbf{v}_1[t]+\mathbf{n}_1[t] & \cdots & (\tilde {\mathbf {H}}_{d}+\tilde {\mathbf {H}}_{2}\mathbf {Q}\tilde {\mathbf {H}}_{1})\mathbf{v}_K[t]+\mathbf{n}_K[t]
		\end{bmatrix}\\
		=&(\tilde{\mathbf {H}}_{d}+\tilde{\mathbf {H}}_{2}\mathbf{Q}\tilde{\mathbf {H}}_{1})\mathbf{V}[t]+\mathbf{N}[t] \label{eq:receivedsignal_Q} ,
	\end{align}
	where $\mathbf{N}[t]=[\mathbf{n}_1[t], \cdots, \mathbf{n}_K[t]]\in\mathbb{C}^{N_r\times K}$ is the complex Gaussian noise matrix, and $\mathbf{V}[t]=[\mathbf{v}_1[t],\cdots,\mathbf{v}_K[t]]$ is defined in (\ref{eq:Vt}), with $\mathbf{v}_k[t]=\mathbf{e}_i[t]v_k[t]$. The relationship between the received vector $\mathbf{y}_k[t]$ and the transmitted vector $\mathbf{v}_k[t]$, at the $k$-th time slot of the $t$-th block, is 
	\begin{align}
		\mathbf {y}_{k}[t]=(\tilde {\mathbf {H}}_{d}+\tilde {\mathbf {H}}_{2}\mathbf {Q}\tilde {\mathbf {H}}_{1})\mathbf {v}_{k}[t]+\mathbf {n}_{k}[t], \label{eq:ykt}
	\end{align}
	which is another form of (\ref{eq:recivedvector}). The similarity becomes clear when substituting $\mathbf{v}_k[t]=\mathbf{e}_i[t]v_k[t]$ into (\ref{eq:ykt}), yielding
	\begin{align}
		(\tilde {\mathbf {H}}_{d}+\tilde {\mathbf {H}}_{2}\mathbf {Q}\tilde {\mathbf {H}}_{1})\mathbf {v}_{k}[t]=(\tilde {\mathbf {H}}_{d}+\tilde {\mathbf {H}}_{2}\mathbf {Q}\tilde {\mathbf {H}}_{1})\mathbf{e}_i[t]v_k[t], \label{eq:deriveHv}
	\end{align}
	It can be shown that multiplying $\mathbf{H}=\tilde {\mathbf {H}}_{d}+\tilde {\mathbf {H}}_{2}\mathbf {Q}\tilde {\mathbf {H}}_{1}\in\mathbb{C}^{N_r\times K}$ by $\mathbf{e}_i[t]\in\mathbb{C}^{K\times1}$ results in a vector in $\mathbb{C}^{N_r\times1}$, and each $v_k[t]$ directly multiplies with the $i$-th column vector of the matrix $\mathbf{H}$ in (\ref{eq:Htotal}). Each column vector is defined as $\mathbf{h}_d+\mathbf{H}_2\mathbf{\Phi}_i\mathbf{h}_1$ in (\ref{eq:H}). Thus, (\ref{eq:deriveHv}) can be put in the form
	\begin{align}
		(\tilde {\mathbf {H}}_{d}+\tilde {\mathbf {H}}_{2}\mathbf {Q}\tilde {\mathbf {H}}_{1})\mathbf{e}_i[t]v_k[t]=
		(\mathbf{h}_d+\mathbf{H}_2\mathbf{\Phi}_i\mathbf{h}_1)v_k[t]\in\mathbb{C}^{N_r\times 1},
	\end{align}
	since $\mathbf{h}_d\in\mathbb{C}^{N_r\times 1}$ and $\mathbf{H}_2\mathbf{\Phi}_i\mathbf{h}_1\in\mathbb{C}^{N_r\times 1}$.
	Hence, (\ref{eq:recivedvector}) and (\ref{eq:ykt}) are equivalent to each other, which is
	\begin{align}
		\mathbf {y}_{k}[t]=(\tilde {\mathbf {H}}_{d}+\tilde {\mathbf {H}}_{2}\mathbf {Q}\tilde {\mathbf {H}}_{1})\mathbf {v}_{k}[t]+\mathbf {n}_{k}[t]=(\mathbf{h}_d+\mathbf{H}_2\mathbf{\Phi}_i\mathbf{h}_1)v_k[t]+\mathbf {n}_{k}[t].
	\end{align}
	From (\ref{eq:receivedsignal_Q}) and (\ref{eq:ykt}) we finally have
	\begin{align}
		\mathbf {y}_{k}[t]&=\mathbf {H}\mathbf {v}_{k}[t]+\mathbf {n}_{k}[t], \\
		\mathbf {Y}[t]&=\mathbf {H}\mathbf {V}[t]+\mathbf {N}[t]. \label{eq:receivedsignal}
	\end{align}
	
	By substituting (\ref{eq:diffencode}) into (\ref{eq:receivedsignal}), the received signal is
	\begin{align}
		\mathbf{Y}[t]=\mathbf{H}\mathbf{V}[t-1]\mathbf{X}[t]+\mathbf{N}[t]=\mathbf{Y}[t-1]\mathbf{X}[t]-\mathbf{N}[t-1]\mathbf{X}[t]+\mathbf{N}[t]. \label{eq:deriveYt}
	\end{align}
	With $\tilde{\mathbf{H}}_{d}$, $\tilde{\mathbf{H}}_{2}$, and $\tilde{\mathbf{H}}_{1}$ being unknown at the receiver,  maximum likelihood (ML) detection aims to detect
	 $\hat{\mathbf{X}}[t]$ such that the likelihood of receiving $\mathbf{Y}[t]$ is maximum, given $\mathbf{Y}[t-1]$.
	
	Therefore, the optimal ML detection rule can be written by \cite{MLdetect,MLdetect2,DRM}
	\begin{align}
		\hat {\mathbf {X}}[t]&=\arg \min \limits_{\mathbf {X}[t]\in \mathcal {X}}\lVert\mathbf {Y}[t]-\mathbf {Y}[t-1]\mathbf {X}[t]\rVert_{F}^{2}, \label{eq:MLdetect} \\
		&=\arg \max \limits_{\mathbf {X}[t]\in \mathcal {X}} \Re \left \{{\rm {tr}}(\mathbf {Y}[t]\mathbf {Y}^{H}[t-1]\mathbf {X}[t])\right \}, \label{eq:MLdetection}
	\end{align}
	where $\mathcal{X}$ is the set of all legitimate $\mathbf{X}[t]$ and $\lvert\mathcal{X}\rvert=2^r$. This is recognized as conventional differential detection (CDD) in matrix form not requiring the channel matrix $\mathbf{H}$ from (\ref{eq:receivedsignal}).  The derivation of (\ref{eq:MLdetection}) is provided in Appendix \ref{apdx2}.
	
	\subsubsection{Reflecting Pattern Selection}\label{section:RPS}
	Since the first block does not carry any information, the transmission rate for DRM is given by
	\begin{align}
		R=\frac{(T-1)\ {\rm{blocks}}\times r\ {\rm{bits/block}}}{T\ {\rm {blocks}}\times K\ {\rm {symbols/block}}}=\frac {(T-1)(\lfloor {\rm log}_2K!\rfloor+K\log_2M)}{TK}, 
	\end{align}
	where $T$ is the total number of blocks. It can be easily confirmed that the rate increases gradually with $K$, as $M$ is held constant \cite{JiaweiThesis}. The computational complexity of detection can be estimated as $2^r(K^2N_r + K^3)$ multiplications \cite{DRM}. This is because $\mathbf{Y}^H[t] \mathbf{Y}[t-1] \mathbf{X}[t]$ must be computed $|\mathcal{X}| = 2^r$ times to test all possible values of $\mathbf{X}[t] \in \mathcal{X}$. Each test requires $K^2N_r + K^3$ multiplications. Therefore, the complexity can be expressed as \cite{DRM}
	\begin{align}
		C_{uc}=&2^r(K^2N_r+K^3)=2^{\lfloor \log_2K!\rfloor+K\log_2M}(K^2N_r+K^3) \notag \\
		=&(2^{\lfloor \log _{2} K!\rfloor}\cdot M^K)(K^2N_r+K^3).
	\end{align}
	When $N_r$ and $M$ are constant, the detection complexity increases significantly with $K$.
	
	Minimizing the number of reflecting patterns $K$ can reduce complexity. As shown in \cite{RM, tse2005fundamentals}, the bit error rate (BER) for DRM is upper bounded by
	\begin{align} P_{b}\leq \overline {P_{b}}=\frac {1}{2^r}\sum _{a=1}^{2^r}\sum _{\hat {a}=1,\hat {a}\neq a}^{2^r}D_{\mathrm{HD}}(\textbf {b}_{a},\textbf {b}_{\hat {a}})P_{\mathrm{PEP}}(a\rightarrow \hat {a}), \label{eq:Pb}
	\end{align}
	where $a$ denotes the index of the set of transmitted symbols and permutation matrices mapped by bit sequence $\mathbf{b}_a$, $\mathbf{b}_{\hat{a}}$ and $\hat{a}$ are the corresponding detected index and bit sequence by CDD, $D_{\mathrm{HD}}$ is the Hamming distance between $\mathbf{b}_a$ and $\mathbf{b}_{\hat{a}}$, with $P_{\mathrm{PEP}}$ idenoting the pairwise error probability. In (\ref{eq:Pb}), $P_{\mathrm{PEP}}(a\rightarrow \hat {a})$ is expressed as \cite{RM}
	\begin{align} P_{\mathrm{PEP}}(a\rightarrow \hat {a})=Q\left ({\sqrt {\frac {D_{\mathrm{ED}}(a,\hat {a})^{2}}{2\sigma ^{2}}}}\right)\!,\label{eq:PEP}
	\end{align}
	where 
	\begin{align}
		D_{\mathrm{ED}}(a,\hat {a})^{2}\!=\!\lVert(\textbf {h}_{d}+\textbf {H}_{2}\boldsymbol{\Phi }_{k}\textbf {h}_{1})s_k-(\textbf {h}_{d}\!+\!\textbf {H}_{2}\boldsymbol{\Phi }_{k'}\textbf {h}_{1}){s}_{k'}\rVert_{2}^{2},\!\!\!\! \label{eq:ED}
	\end{align}
	Here, $D_{\mathrm{ED}}$ is the Euclidean distance between two received signal symbols,  $\lVert\cdot\rVert_2$ denotes the 2-norm, and $Q(\cdot)$ is the Q-function associated with the Gaussian distribution. Thus, both Hamming and Euclidean distances influence the BER.
	
	Lemma 1 in \cite{lemma1} indicates that at high SNR, the upper bound $\overline{P_b}$ is more sensitive to changes in Euclidean distances than Hamming distances. When SNR is sufficiently large, the weighted pairwise error probability $P_{w, \mathrm{PEP}}(a \to \hat{a}) = D_{\mathrm{HD}} P_{\mathrm{PEP}}(a \to \hat{a})$ can be approximated as \cite{tse2005fundamentals}
	\begin{align} P_{w,\mathrm{PEP}}(a\rightarrow \hat {a}) \approx D_{\mathrm{HD}}(\textbf {b}_{a},\textbf {b}_{\hat {a}}) \begin{pmatrix}2N_{r}-1\\N_{r}-1\end{pmatrix}\left[{\rho D_{\mathrm{ED}}(a,\hat {a})}\right ]^{-N_{r}}, \end{align}
	where $\rho=1/\sigma^2$ is the SNR, and $\begin{pmatrix}m\\n\end{pmatrix}$ is the binomial coefficient.
	
	Consequently, the optimization focuses on selecting the signal mapping and reflecting patterns that maximize $D_{\mathrm{ED}}(a, \hat{a})$ to reduce complexity. Without CSI, as $\mathbf{h}_d$, $\mathbf{H}_2$, and $\mathbf{h}_1$ are fixed within a frame, the optimization criterion proposed in \cite{DRM, RM} maximizes the minimum mutual Euclidean distances
	\begin{align}
		& D_{ED,\min}=\min_{\substack{\boldsymbol{\Phi }_{i},\boldsymbol{\Phi }_{i'}\in \Psi,s_{k},s_{k'}\in \mathcal {S}_{M}\\ {\boldsymbol{\Phi }_{i}s_{k}\neq \boldsymbol{\Phi }_{i'} s_{k'}}}}\lVert\boldsymbol{\Phi }_{i}s_{k}-\boldsymbol{\Phi }_{i'}s_{k'}\rVert_{2}, \\
		& \mathrm{Maximize}\!: D_{ED,min}, \notag
	\end{align}
	The stepwise depletion algorithm described in \cite{RM} identifies the optimized set of $K$ reflecting patterns $\boldsymbol{\Phi}_i \in \Psi$, based on the known $M$-PSK constellation $\mathcal{S}_M$ when CSI is unknown at the transmitter, receiver, and RIS. The stepwise depletion algorithm starts by initializing all possible candidates from the legitimate tuples $(\mathbf{\Phi}_i, s_k)$, where $\mathbf{\Phi}_i \in \Psi$ and $s_k \in \mathcal{S}_M$. It gradually reduces the $2^N$ legitimate patterns to $K$. The algorithm begins by selecting one tuple, calculating the Euclidean distances between any two of the remaining tuples, recording the minimal value, and removing the tuple with the smallest recorded Euclidean distance. This process repeats until only $K$ legitimate tuples remain, which are then assembled into $\mathbf{Q}$ diagonally as in (\ref{eq:Q}).
	
	\subsection{DRM-DSTM Basics}\label{section:DSTM}
	Space-time block codes with a group structure were first introduced in \cite{DSTM}. The system involves $N_t$ transmit antennas and use a constellation $\mathcal{C}$ based on $M$-PSK. During the encoding process, $K$ symbols are transmitted. A space-time code is a set of symbols from the constellation $\mathcal{C}$, represented as a matrix $\mathbf{C}\in\mathcal{C}^{N_t\times K}$. The code is considered unitary if its codewords are square matrices that satisfy $\mathbf{C}\mathbf{C}^{\rm H}=\mathbf{I}$. For $K\geq N_t$, let $\mathcal{G}$ be a group of matrices where, for any $\mathbf{G}\in\mathcal{G}$, we have $\mathbf{G}^{\rm H}\mathbf{G}=\mathbf{G}\mathbf{G}^{\rm H}=\mathbf{I}$ and $\mathbf{G}\in\mathbb{C}^{K\times K}$. Let $\mathbf{D}$ be a $N_t\times K$ matrix such that for any $\mathbf{G}\in\mathcal{G}$, $\mathbf{D}\mathbf{G}$ belongs to $\mathcal{C}^{N_t\times K}$. The set $\mathbf{D}\mathcal{G}=\{\mathbf{D}\mathbf{G}:\mathbf{G}\in\mathcal{G}\}$ forms a group code of length $K$, with the transmission rate $R=(1/K)\log_2|\mathcal{G}|$ bits/s/Hz, where $|\mathcal{G}|$ is the cardinality of the group. Since there are $|\mathcal{G}|$ code matrices, the number of information bits is $r_2'=\log_2|\mathcal{G}|$, and the rate is the ratio of information bits to the code length.
	
	\begin{figure}[!htbp]
		\centering
		\includegraphics[width=0.8\textwidth]{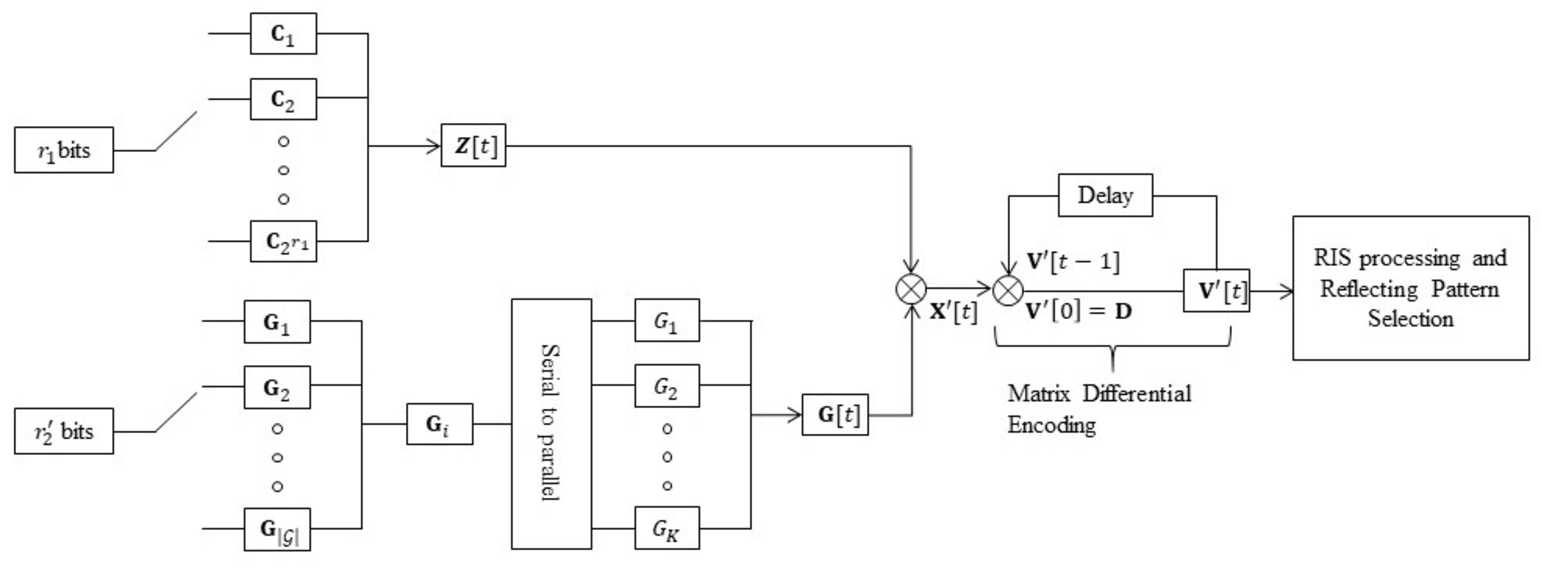}
		\caption{DRM-DSTM encoding.}
		\label{DSTM2DRM}
	\end{figure}
	
	The key idea behind DSTM is that group codes can be differentially encoded similarly to PSK, and hence they can be integrated into DRM. The encoding structure for DRM-DSTM  is illustrated in Fig. \ref{DSTM2DRM}. Information bits are mapped to $\mathbf{G}_i\in\mathcal{G}, i=1,2,\dots,|\mathcal{G}|$. Instead of using the identity matrix for initialization as in the DRM scheme, the transmitter sends $\mathbf{V}'[0]=\mathbf{D}$. In (\ref{eq:InforCarry}), $\mathbf{X}[t]$ is the product of two square matrices, and both $\mathbf{X}[t]$ and $\mathbf{V}[t]$ are $K\times K$. Thus, for the DRM- DSTM scheme, $\mathbf{V}'[0]$ must also be a $K\times K$ matrix. In DRM-DSTM, the dimensions of $\mathbf{D}\mathbf{G}$ depend on $N_t$ and $K$, and $\mathbf{G}$ is always a $K\times K$ matrix. Therefore, $\mathbf{D}$ is chosen as a $N_t\times K$ matrix over $\mathcal{C}$. Only when $N_t=K$ will $\mathbf{G}$ be a square matrix. By replacing $\mathbf{S}[t]$ with $\mathbf{G}[t]\in\mathcal{G}$ in (\ref{eq:InforCarry}), a space-time code can be applied to the DRM system. This is achieved by connecting the $N_t$ transmit antennas in the DSTM system to the $K$ symbol slots during the DRM encoding process, as depicted in Fig. \ref{DSTM2DRM}. In this figure, $\mathbf{G}_i\in\mathcal{G}$ is determined by the information bits $r_2'$, and it has the same structure as a permutation matrix, with one non-zero element in each column \cite{DSTM}. Let $G_{k}$ represent the element in the $k$-th column of $\mathbf{G}_i$, which is transmitted by $N_t$ antennas and then assembled into the matrix $\mathbf{G}[t]=\mathbf{G}_i$ for the $t$-th block in a frame. The transmitted messages are differentially encoded, with the transmitter sending
	\begin{align}
		\mathbf{V}'[t]=\mathbf{V}'[t-1]\mathbf{X}'[t], \label{eq:DSTMencode}
	\end{align}
	where $\mathbf{X}'[t]=\mathbf{Z}[t]\mathbf{G}[t]$. The ML decoder for the differentially encoded unitary matrix $\mathbf{G}[t]$, as shown in \cite{DSTM}, is
	\begin{align}
		\hat{\mathbf{G}}[t]=\arg\max\limits_{\mathbf {G}[t]\in \mathcal {G}} \Re \left \{{\rm{tr}}(\mathbf{Y}^{H}[t]\mathbf{Y}[t-1]\mathbf{G}[t])\right \}, 
	\end{align}
	which is derived in the same way as the CDD for $\hat{\mathbf{X}}[t]$. Therefore,  ML detection for the coded system can be expressed as
	\begin{align}
		\hat{\mathbf{X}}'[t]=\arg\max\limits_{\mathbf {X}'[t]\in \mathcal {X}'} \Re \left \{{\rm{tr}}(\mathbf{Y}^{H}[t]\mathbf{Y}[t-1]\mathbf{X}'[t])\right \}, \label{eq:DSTMMLdetect}
	\end{align}
	where $\mathcal{X}'$ denotes the set of all legitimate $\mathbf{X}'[t]$, with cardinality $|\mathcal{X}'|=2^{r_1+r_2'}$.
	
	\subsection{Performance of DRM  and DRM-DSTM Systems for $N_t=K=2$}\label{section:2.3}
	
	In this section, we present  simulation results for small configurations DRM and DRM-DSTM with $N_t=K=2$, using BPSK and QPSK symbols. Firstly, to evaluate the performance of the uncoded system, we compare DRM  with a MIMO non-DRM (NDRM) scheme where  the transmitted signal $\mathbf{X}[t]$ is received as $\mathbf{Y}'[t]=\mathbf{H}\mathbf{X}[t]+\mathbf{N}[t]$. For NDRM, perfect CSI is assumed available  for coherent detection, and the decoding process is given by
	\begin{align}
		\hat{\mathbf{X}}[t]=\arg\min\limits_{\mathbf{X}[t]\in\mathcal{X}}{\rm Tr}\{(\mathbf{Y}'[t]-\mathbf{H}\mathbf{X}[t])(\mathbf{Y}'[t]-\mathbf{H}\mathbf{X}[t])^{H}\}. \label{eq:nondiffdecode}
	\end{align}
	As in \cite{DRM}, the number of receive antennas $N_r$ is 3, and the RIS is assumed to have $N=4$ units, with diagonal elements $(\mathbf{\Phi}_k)_{nn}\in\{-1,1\}$, resulting in $2^N=16$ possible reflecting patterns. Specific reflecting patterns used in the simulation are selected according to section \ref{section:RPS}.
	
	In the simulation, the SNR is defined as $\rho=\frac{E\{|v_k|^2\}}{\sigma^2}=\frac{1}{\sigma^2}$, where $\sigma^2$ denotes the variance of each component in $\mathbf{n}_k[t]$.  For each SNR, 10000 iterations are conducted using Monte Carlo simulations over the uncorrelated quasi-static Rayleigh fading channels, with $10^2$ blocks of information bits randomly generated following a discrete uniform distribution. As mentioned in section \ref{section:Encoding}, $\mathbf{h}_d$, $\mathbf{h}_1$, $\mathbf{H}_2$ are modeled as uncorrelated quasi-static Rayleigh fading channels, where the fading coefficient $h$ is zero-mean complex Gaussian with normalized power $E\{|h|^2\}=1$. The BER is computed for each SNR.

	\begin{figure}[htbp!]
		\centering
		\includegraphics[width=0.5\textwidth]{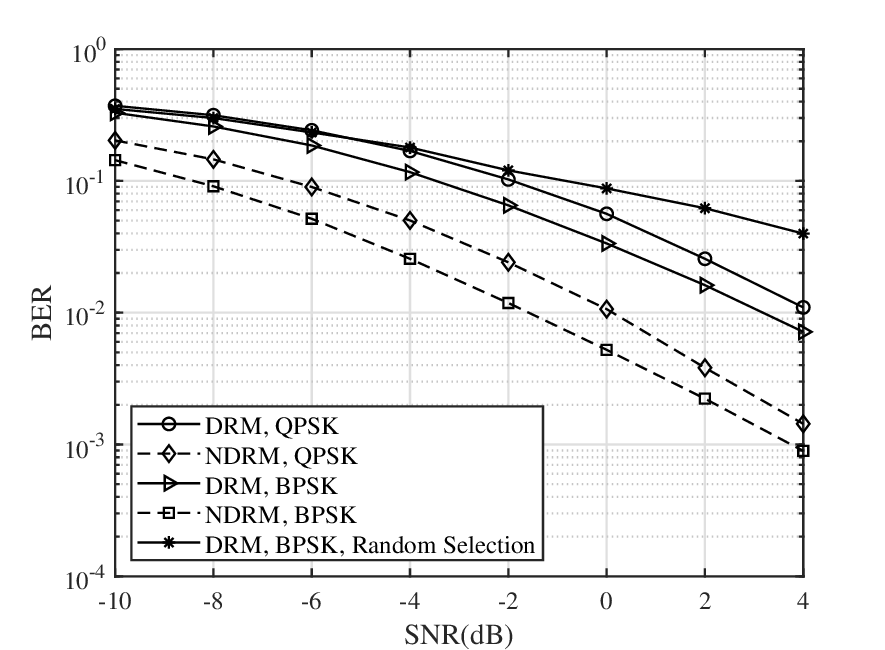}
		\caption{Performance of DRM in a low SNR range.}
		\label{rep_frob}
	\end{figure}
	
	Firstly, we present in Fig. \ref{rep_frob} a comparison between DRM and NDRM. The DRM scheme shows inferior performance compared to NDRM. Specifically, for BPSK, NDRM is better than DRM by 4.8 dB at a BER of $10^{-2}$, while for QPSK, the difference is 4.1 dB. This performance gap is expected, given that DRM does not require CSI. Moreover, it is clear that BPSK consistently performs better than QPSK in both DRM and NDRM schemes. For example, in the DRM scheme, BPSK outperforms QPSK by around 1 dB at a BER of $10^{-2}$. Additionally, the effectiveness of the proposed reflecting pattern selection method, as discussed in section \ref{section:RPS}, is demonstrated when compared to a random pattern selection  $\mathbf{Q}_{\rm RD}={\rm diag}(1,-1,-1,-1,-1,-1,-1,-1)$. As illustrated in Fig. \ref{rep_frob}, for the DRM scheme with BPSK, at $E_b/N_0 = 4$ dB, the optimized reflecting patterns yield a BER of approximately $7\cdot10^{-3}$, while a randomly chosen pattern results in a significantly worse BER of around $4\cdot10^{-2}$. These observations correlate well with those from \cite{DRM}.

	Next we expand our analysis to include also a DRM- DSTM-coded scheme. We simulate up to $r\cdot10^{9}$ information bits and gather at least 30,000-bit errors for the NDRM scheme and 100,000-bit errors for the DRM and DRM-DSTM  schemes.
We use space-time coding to the DRM-DSTM scheme and compare its BER performance to original DRM.
Following \cite{DSTM}, $\mathbf{D}$ can be any matrix that satisfies $\mathbf{D}\mathbf{D}^{\rm H}=2\mathbf{I}$, with $\setlength{\arraycolsep}{1.5pt}\mathbf{D}=\begin{bmatrix}1&-1\\[-6pt]1&1\end{bmatrix}$ being a simple choice. The notation $\langle\cdot\rangle$ denotes the group generated by the enclosed elements. For $M$-PSK, the cyclic group code is defined as $\mathcal{G}_c=\langle \mathbf{G}_1\rangle=\{\mathbf{I},\mathbf{G}_1,\cdots,\mathbf{G}_1^{M-1}\}$, while the dicyclic group code is defined as $\mathcal{G}_{dc}=\langle \mathbf{G}_1,\mathbf{G}_2\rangle=\{\mathbf{G}_1^m\mathbf{G}_2^l:0\leq m<\frac{M}{2};l=0,1\}$. The unitary group codes $\mathcal{G}$ used in the simulations for different constellations are shown in Table \ref{tb:groupK2}.

	\begin{table}[!htb]
		\centering
		\caption{Unitary group code for $K=2$}
		\setlength{\arraycolsep}{1pt}
		\begin{tabular}{c|c}
			\hline
			Constellation $\mathcal{C}$ & Unitary group code $\mathcal{G}$\\
			\hline
			BPSK & $\langle \left[ \begin{matrix} -1 & 0 \\[-3pt] 0 & -1 \end{matrix}\right] \rangle$\\
			QPSK & $\langle \left[ \begin{matrix} j & 0\\[-3pt] 0&-j \end{matrix}\right], \left[ \begin{matrix} 0&-1\\[-3pt] 1&0 \end{matrix} \right] \rangle$\\
			\hline
		\end{tabular}
		\label{tb:groupK2}
	\end{table}

	\begin{figure}[htbp!]
		\centering
		\includegraphics[width=0.5\textwidth]{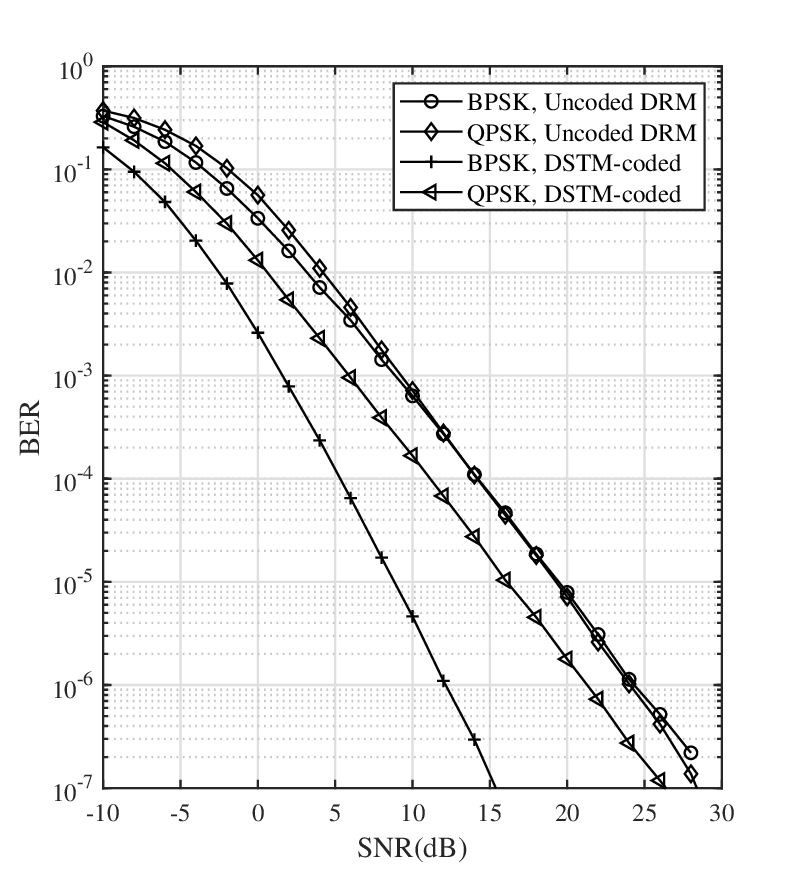}
		\caption{Performance of coded DRM-DSTM  compared with the uncoded systems for $K=2$.}
		\label{CodedvsUncoded}
	\end{figure}

	In Fig. \ref{CodedvsUncoded}, we compare the performance of the uncoded DRM  with the DRM-DSTM short coded system of $K=2$ specified in Table \ref{tb:groupK2}. For both BPSK and QPSK, the coded system demonstrates significantly better error performance than the original uncoded  DRM scheme. At a BER of $10^{-6}$, DRM-DSTM improves performance by roughly 12 dB for BPSK and around 3 dB for QPSK.  However, t Furthermore, the DSTM-coded scheme with BPSK outperforms the one with QPSK by 10 dB at a BER of $10^{-6}$.
	
	\section{General DRM-DSTM  based on codes  with a Group Structure}\label{section3}
	In this section, we examine space-time cyclic and dicyclic codes with a group structure for DRM-DSTM. We provide tables of suitable unitary group codes, and consider the computational complexity for both uncoded DRM  and DRM-DSTM coded systems.
	
	\subsection{ Construction of Coded Systems}\label{section3.1}
	We further investigate the  properties of DRM-DSTM  with a group structure to extend the coded system to a general format suitable for a variety of systerm parameters. In section \ref{section:DSTM}, the size of the space-time codes used in DRM-DSTM  has been  limited to $N_t=K=2$. In \cite{DSTM}, $\mathcal{G}_c$ is defined as a cyclic group code $(M;u_1)$ where $u_1$ is an odd integer parameter with $0<u_1<M$, and $|\mathcal{G}_c|=M$. The symbols in the code are based on  $M$-PSK , while $\mathcal{G}_{dc}$ denotes a dicyclic group code with symbols in a $\frac{M}{2}$-PSK constellation and $|\mathcal{G}_{dc}|=M$. The codes we considered in the previous section and specified in Table \ref{tb:groupK2} are defined by
	\begin{align}
		\mathcal{G}_c &= \bigg\langle \left[ \begin{matrix} -1 & 0 \\[-3pt] 0 & -1 \end{matrix}\right] \bigg\rangle \triangleq \bigg\langle \left[ \begin{matrix} e^{\frac{2\pi j}{M}} & 0 \\[-3pt] 0 & e^{\frac{2\pi j}{M}u_1} \end{matrix}\right] \bigg\rangle \label{eq:cycgrp1}\\
		\mathcal{G}_{dc} &= \bigg\langle \left[ \begin{matrix} j & 0\\[-3pt] 0&-j \end{matrix}\right], \left[ \begin{matrix} 0&-1\\[-3pt] 1&0 \end{matrix} \right] \bigg\rangle \triangleq  \bigg\langle \left[ \begin{matrix} e^{\frac{2\pi j}{M/2}} & 0\\[-3pt] 0&e^{(\frac{2\pi j}{M/2})^*} \end{matrix}\right], \left[ \begin{matrix} 0&-1\\[-3pt] 1&0 \end{matrix} \right] \bigg\rangle, \label{eq:dicgrp1}
	\end{align}
	where $M=2$, and $u_1=1$. Let $o$ represent the maximum order of any element in a group $\mathcal{G}$. According to \cite{DSTM}, unitary group codes with $|\mathcal{G}|=M$ and $N_t=K=2$ are either cyclic if $o=M$ or dicyclic if $o=\frac{M}{2}$. A cyclic group code is given by $\mathcal{G}_c=\langle \mathbf{G}_1\rangle=\{\mathbf{I},\mathbf{G}_1,\cdots,\mathbf{G}_1^{M-1}\}$, while a dicyclic group code is $\mathcal{G}_{dc}=\langle \mathbf{G}_1,\mathbf{G}_2\rangle=\{\mathbf{G}_1^m\mathbf{G}_2^l:0\leq m<\frac{M}{2};l=0,1\}$.
	For example, over QPSK constellation, \begin{align}\setlength{\arraycolsep}{1.5pt}\mathcal{G}=\left\{ \pm\left[ \begin{matrix} 1 & 0 \\[-6pt] 0 & 1 \end{matrix}\right], \pm\left[ \begin{matrix} j & 0 \\[-6pt] 0 & -j \end{matrix}\right], \pm\left[ \begin{matrix} 0 & 1 \\[-6pt] -1 & 0 \end{matrix}\right], \pm\left[ \begin{matrix} 0 & j \\[-6pt] j & 0 \end{matrix}\right] \right\}, \end{align} with $|\mathcal{G}|/2=4=M$. In many cases, cyclic group codes with $M$-PSK and dicyclic group codes with $M/2$-PSK represent different constellations, but there are exceptions, such as when both codes use QPSK. In this instance, cyclic codes are based on a $4$-PSK constellation, while dicyclic codes use $8/2$-PSK, corresponding to $M$ values of 4 and 8, respectively. Despite sharing the same constellation, their $M$ values differ.
	
	To make (\ref{eq:cycgrp1}) and (\ref{eq:dicgrp1}) more general, we first rewrite the previous $(M;u_1)$ cyclic group code as $(M;u_1,u_2)$, where $0<u_1\leq u_2<M$ are odd integers, and
	\begin{align}\setlength{\arraycolsep}{1pt}
		\mathcal{G}_{c}=\bigg\langle \left[ \begin{matrix} e^{\frac{2\pi j}{M}u_1} & 0 \\ 0 & e^{\frac{2\pi j}{M}u_2} \end{matrix}\right] \bigg\rangle. \label{eq:cycgrp2}
	\end{align}
	If $u_1=u_2=1$, (\ref{eq:cycgrp2}) is exactly the group code for BPSK in Table \ref{tb:groupK2}. Similarly, we define (\ref{eq:dicgrp1}) as a $(M;u_1)$ dicyclic group code for odd $0<u_1<\frac{M}{2}$, where
	\begin{align}\setlength{\arraycolsep}{1pt}
		\mathcal{G}_{dc}=\bigg\langle \left[ \begin{matrix} e^{\frac{2\pi j}{M/2}u_1} & 0 \\ 0 & e^{\frac{2\pi j}{M/2}u_2} \end{matrix}\right], \left[ \begin{matrix} 0 & -1 \\ 1 & 0 \end{matrix}\right]\bigg\rangle, \label{eq:dicgrp2}
	\end{align}
	and $u_2=-u_1$. By setting $u_1=1$, (\ref{eq:dicgrp2}) is also the same as (\ref{eq:dicgrp1}).
	
	So far, we have presented  general expressions for cyclic and dicyclic group codes when $N_t=K=2$. To broaden the applicability of the proposed DRM-DSTM coded schemes, we extend the formulations in (\ref{eq:cycgrp1}) and (\ref{eq:dicgrp1}) to $N_t=K>2$, following the methods in \cite{Further,Optimal}. 
	
	\textit{Theorem 1} from \cite{Optimal} states that if a $N_t\times K$ group code $\mathbf{D}\mathcal{G}$ is unitary, then $\mathbf{D}\mathbf{D}^{\rm H}=K\mathbf{I}$. For any square $\mathbf{D}$, $N_t=K$, a convenient option is $\mathbf{D}=\sqrt{K}\mathbf{I}_K$, though other alternatives may be available or more optimal \cite{Further}. For example, the matrix $\setlength{\arraycolsep}{1.5pt}\mathbf{D}=\begin{bmatrix}1&-1\\[-6pt]1&1\end{bmatrix}$ used in the previous section is also a valid candidate. The symbols in $\mathbf{D}\mathcal{G}$ take values from a specific constellation, such as $M$-PSK, assisted by $\mathbf{D}$. It can be verified that with (\ref{eq:cycgrp2}) and (\ref{eq:dicgrp2}), both $\setlength{\arraycolsep}{1.5pt}\mathbf{D}=\begin{bmatrix}1&-1\\[-6pt]1&1\end{bmatrix}$ and $\setlength{\arraycolsep}{1.5pt}\mathbf{D}=\sqrt{2}\begin{bmatrix}1&0\\[-6pt]0&1\end{bmatrix}$ can produce appropriate group codes $\mathbf{D}\mathcal{G}$ for $8$-PSK when $(u_1,u_2)=(1,3)$. For example, $\setlength{\arraycolsep}{1.5pt}\mathbf{G}=\begin{bmatrix}e^{\frac{\pi}{4}j}&0\\[-6pt]0&e^{\frac{3}{4}\pi j}\end{bmatrix}$ is a matrix from a $(8;1,3)$ cyclic group code defined by (\ref{eq:cycgrp2}). Applying $\mathbf{D}$ to this matrix, we have
	\begin{align}
		\mathbf{D}\mathbf{G}=& \begin{bmatrix}1&-1\\1&1\end{bmatrix}\begin{bmatrix}e^{\frac{\pi}{4}j}&0\\0&e^{\frac{3}{4}\pi j}\end{bmatrix} \notag\\
		=&\begin{bmatrix}e^{\frac{\pi}{4}j}&-e^{\frac{3}{4}\pi j}\\e^{\frac{\pi}{4}j}&e^{\frac{3}{4}\pi j}\end{bmatrix}=\begin{bmatrix}e^{\frac{\pi}{4}j}&e^{-\frac{\pi}{4} j}\\e^{\frac{\pi}{4}j}&e^{\frac{3}{4}\pi j}\end{bmatrix}
	\end{align}
	where $e^{\frac{\pi}{4}j}$, $e^{-\frac{\pi}{4}\ j}$, and $e^{\frac{3}{4}\pi j}$ are symbols from an 8-PSK constellation.
	
	Therefore, both types of $\mathbf{D}$ are considered. Notably, the matrix $\setlength{\arraycolsep}{1.5pt}\begin{bmatrix}1&-1\\[-6pt]1&1\end{bmatrix}$ is a $2 \times 2$ Hadamard matrix, which only exist when their dimensions are powers of 2. Consequently, the Hadamard matrix is used when $N_t=K$ is a power of 2, while $\sqrt{K}\mathbf{I}_K$ is used in other cases. For instance, when $N_t=K=4$, the Hadamard matrix is
	\begin{align}
		\mathbf{D}=\left[\begin{matrix}
			1&-1&-1&1\\
			1&1&-1&-1\\
			1&-1&1&-1\\
			1&1&1&1
		\end{matrix}\right]. \label{eq:Hadamard4}
	\end{align}
	
	Next, we present general expressions for any $M$-ary space-time unitary group codes with $N_t=K$. Following  (\ref{eq:cycgrp1}), (\ref{eq:dicgrp1}) from \cite{DSTM} and our  (\ref{eq:cycgrp2}), (\ref{eq:dicgrp2}), we can extend both cyclic and dicyclic group codes to $N_t=K>2$. For an $M$-ary constellation and any odd $0<u_1\leq\cdots\leq u_K<M$, an $(M;u_1,\cdots,u_K)$ a cyclic group code is defined as $\mathcal{G}=\langle \mathbf{G}_0 \rangle$,  where
	\begin{align}
		\mathbf{G}_0=\left[\begin{matrix}
			e^{\frac{2\pi j}{M}u_1}&0&\cdots&0\\
			0&e^{\frac{2\pi j}{M}u_2}&\cdots&0\\
			\vdots&\vdots&\ddots&\vdots\\
			0&0&\cdots&e^{\frac{2\pi j}{M}u_K}
		\end{matrix}\right].\label{eq:cyclic}
	\end{align}
	This group code consists of $|\mathcal{G}|=M$ valid matrices with  $M$-PSK symbols when $\mathbf{D}$ is selected as either a Hadamard matrix or $\sqrt{K}\mathbf{I}_K$.
	
	Furthermore, for even $N_t=K$,, and odd integers $0<u_1\leq\cdots\leq u_{K/2}<M/2$, a $(M;u_1,\cdots,u_{K/2})$ dicyclic group code can be represented by $\mathcal{G}=\langle \mathbf{G}_0,\mathbf{G}_1\rangle$, where
	\begin{align}
		\mathbf{G}_0&=\left[\begin{matrix}
			e^{\frac{2\pi j}{M/2}u_1}&0&\cdots&0\\
			0&e^{\frac{2\pi j}{M/2}u_2}&\cdots&0\\
			\vdots&\vdots&\ddots&\vdots\\
			0&0&\cdots&e^{\frac{2\pi j}{M/2}u_K}
		\end{matrix}\right]\label{eq:dicyclicG0} \\
		\mathbf{G}_1&=\left[\begin{matrix}
			0&-\mathbf{I}_{K/2}\\
			\mathbf{I}_{K/2}&0
		\end{matrix}\right],\label{eq:dicyclicG1}
	\end{align}
	with $u_{i+\frac{K}{2}}=-u_i,i=1,\cdots,\frac{K}{2}$. This code consists of $\frac{M}{2}$-PSK symbols and has a cardinality of $|\mathcal{G}|=M$.
	
	As previously mentioned, \cite{DSTM} indicates that all unitary group codes with $|\mathcal{G}|=M$ are equivalent to either cyclic or dicyclic codes for $N_t= K=2$. Moreover, \textit{Theorem 2} in \cite{Further} indicates the same property for $N_t=K>2$. Specifically, for odd $N_t=K$, a unitary group code with $M$ matrices is equivalent only to an $(M;u_1,\cdots,u_K)$ cyclic code. For even $N_t=K$, on the other hand, the code can be either  cyclic  $(M;u_1,\cdots,u_K)$or  dicyclic $(M;u_1,\cdots,u_{K/2})$. Without loss of generality, we can set $u_1=1$ and the constellation is then determined by $u_2,\cdots,u_K$. These theorems narrow the range of potential unitary group codes. Subsequently, simulations are conducted to assess the performance of the proposed codes in our DRM-DSTM coded system.
	
	\begin{table}[!htb]
		\begin{minipage}{.5\linewidth}
			\caption{Unitary Group Codes ($N_t=K=2$)}
			\centering
			\setlength{\tabcolsep}{5mm}{
			\begin{tabular}{c|c}
				\hline  {Cyclic group code} & {Dicyclic group code}\\
				\hline $(M;u_1,\cdots,u_{K})$  & $(M;u_1,\cdots,u_{K/2})$ \\
				\hline 
				$(2;1,1)$ & $(4;1)$ \\
				$(4;1,1)$ & $(8;1)$ \\
				$(8;1,3)$ & $(16;1)$ \\
				$(16;1,7)$ & $(32;1)$ \\
				$(32;1,7)$ & $(64;1)$ \\
				$(64;1,19)$ & $(128;1)$ \\
				$(128;1,47)$ &  \\
				$(256;1,75)$ &  \\
				\hline
		\end{tabular}}
		\label{tb:K=2}
		\end{minipage}%
		\begin{minipage}{.5\linewidth}
			\centering
			\caption{Unitary Group Codes ($N_t=K=3$)}
			\setlength{\tabcolsep}{5mm}{
				\begin{tabular}{c}
					\hline Cyclic group code\\
					\hline $(M;u_1,\cdots,u_{K})$ \\
					\hline 
					$(2;1,1,1)$ \\
					$(4;1,1,1)$ \\
					$(8;1,1,3)$ \\
					$(16;1,3,5)$ \\
					$(32;1,7,9)$ \\
					$(64;1,17,19)$ \\
					\hline
			\end{tabular}}
			\label{tb:K=3}
		\end{minipage}
	\end{table}
	
	\begin{table}[!htb]
		\begin{minipage}{.5\linewidth}
			\caption{Unitary Group Codes ($N_t=K=4$)}
			\centering
			\setlength{\tabcolsep}{5mm}{
				\begin{tabular}{c|c}
					\hline {Cyclic group code} & {Dicyclic group code}\\
					\hline $(M;u_1,\cdots,u_{K})$  & $(M;u_1,\cdots,u_{K/2})$ \\
					\hline 
					$(2;1,1,1,1)$ & $(4;1,1)$ \\
					$(4;1,1,1,1)$ & $(8;1,1)$ \\
					$(8;1,1,3,3)$ & $(16;1,3)$ \\
					$(16;1,3,5,7)$ & $(32;1,7)$ \\
					$(32;1,7,9,15)$ & $(64;1,7)$ \\
					$(64;1,11,17,19)$ & $(128;1,19)$ \\
					$(128;1,29,37,39)$ & \\
					\hline
			\end{tabular}}
			\label{tb:K=4}
		\end{minipage}%
		\end{table}

	The unitary group codes used are presented in  tables \ref{tb:K=2}, \ref{tb:K=3}, \ref{tb:K=4} , based on \cite{Further,Optimal,Represent}. These tables illustrate some of the applicable cyclic or dicyclic group codes, following the definitions in (\ref{eq:cyclic})-(\ref{eq:dicyclicG1}) for various $M$ and $K$. For $N_t=K=2$, both cyclic and dicyclic group codes are applicable, and $\mathbf{D}$ is chosen as a Hadamard matrix, as well as for $N_t=K=4$. However, for $N_t=K=3$ and $5$, only cyclic group codes are valid, and $\mathbf{D} = \sqrt{K} \mathbf{I}_K$ is applied to initialize the code. Additionally, for larger values of $N_t = K$, focusing solely on cyclic group codes does not lead to a noticeable degradation in performance \cite{Optimal}. Different from the approach in \cite{Optimal}, our codes are not restricted by their rate $R$, and we also consider some of the unitary codes introduced in \cite{Represent,Further}.

	\subsection{Complexity Analysis}\label{section: DSTMcomplex}
	In this section, we evaluate the coding and decoding complexity for both  DRM and DRM-DSTM systems. The complexity is typically estimated by counting the number of operations the algorithm performs, such as additions and multiplications, with each elementary operation assumed to take a fixed amount of time.
	
	We begin by analyzing the complexity of the uncoded scheme. During the encoding process, permutation matrices and $M$-PSK symbols are generated. This process has a complexity of $O\left(\binom{K!}{2^{r_1}}\right)$ for selecting $2^{r_1}$ matrices from $K!$ possible candidates, and $O(K\log_2 M)$ for mapping $K$ $M$-PSK symbols. The  complexity of generating the information-carrying matrix in (\ref{eq:InforCarry}) involves $O(K^3)$  matrix multiplications and differential encoding in (\ref{eq:diffencode}). Additionally, generating the channel coefficient matrix $\mathbf{H}=\tilde{\mathbf{H}}_d+\tilde{\mathbf{H}}_2\mathbf{Q}\tilde{\mathbf{H}}_1$ requires $K^2N^2N_r+K^2NN_r$ multiplications and $N_rK$ additions. Subsequently, Signal transmission requires $K^2N_r$ multiplications and $N_rK$ additions, as shown in (\ref{eq:receivedsignal}). Hence, the complexity of these operations is $O(K^2)$.
	
	For the decoding process, as previously discussed, maximum-likelihood (ML) detection involves $K^2N_r+K^3$ multiplications, as indicated in (\ref{eq:MLdetection}). Thus, the  complexity for these multiplications is $O(K^3)$. Furthermore, iterating over $|\mathcal{X}| = 2^r$ possible candidates of $\mathbf{X}[t] \in \mathcal{X}$ leads to a total complexity of $O(2^rK^3)$, which can also be written as $O((2^{\lfloor\log_{2}K! \rfloor}\cdot M^K)K^3)$. It is worth noting that computing the trace, which involves summing $K$ diagonal elements in a $K \times K$ diagonal matrix, requires $K-1$ additions, resulting in a complexity of $O(K)$.
	
	Next, the complexity of the coded system is discussed. Many of the processes have the same results as the uncoded system.  For instance, during the encoding process, generating $\mathbf{Z}[t]$ in (\ref{eq:InforCarry}) also incurs a complexity of $O(\binom{K!}{2^{r_1}})$. Additionally, matrices multiplications in (\ref{eq:DSTMencode}) and $\mathbf{X}'[t]$ require $O(K^3)$ operations. However, different from the previous case where $K$ $M$-PSK symbols were mapped, the complexity of mapping the information bits to the unitary matrix $\mathbf{G}$ is now $O(r_2')$. For cyclic group codes, the bits for $M$ code matrices are $r_2'=\log_{2}M$, while for dicyclic group codes, there are $2M$ code matrices with $r_2'=\log_{2}2M$ bits. The complexity of the modulation remains $O(K^2)$.
	
	As in section \ref{section:RPS}, we consider now the detection  complexity of the coded system. The detection operation $\mathbf{Y}^{H}[t]\mathbf{Y}[t-1]\mathbf{X}'[t]$ in (\ref{eq:DSTMMLdetect}) requires $C_{c}=2^{r'}(K^2N_r+K^3)$ multiplications, where $r'=\lfloor \log_{2}K! \rfloor + r_2'$. For cyclic group codes, $r_2'=\log_{2}M$, while for dicyclic group codes, $r_2'=\log_{2} 2M$. Consequently, for cyclic group codes, the complexity can be written by $C_{c} = (2^{\lfloor \log_{2}K! \rfloor}\cdot M)(K^2N_r+K^3)$, and  $C_{c} = (2^{\lfloor \log_{2}K! \rfloor}\cdot 2M)(K^2N_r+K^3)$ for dicyclic group codes. Hence, the complexity of matrices multiplications and trace calculations during detection, as illustrated in (\ref{eq:DSTMMLdetect}), is $O(K^3)$ and $O(K)$, respectively. With $|\mathcal{X}'|=2^{r'}=2^{r_1+r_2'}$ iterations, the total complexity for the coded scheme becomes  $O(2^{r'}K^3)$.

	\begin{table}[!htbp]
		\centering
		\caption{Summary of  complexity results  of DRM and DRM- DSTM schemes}
		\begin{tabular}{c|c|c|c}
			\hline
			& \multicolumn{3}{c}{Complexity}\\
			\hline
			& \multirow{2}{*}{DRM scheme} & \multicolumn{2}{c}{DRM-DSTM scheme} \\
			\cline{3-4}
			& & Cyclic code& Dicyclic code\\
			\hline
			Permutation matrix & $O(\binom{K!}{2^{r_1}})$ & \multicolumn{2}{c}{$O(\binom{K!}{2^{r_1}})$} \\
			\hline
			Information bits mapping & $O(K\log_2M)$ & $O(\log_{2}M)$ & $O(\log_{2}2M)$\\
			\hline
			Differential Encoding & $O(K^3)$ & \multicolumn{2}{c}{$O(K^3)$}\\
			\hline
			Channel matrix & $O(K^2)$ & \multicolumn{2}{c}{$O(K^2)$}\\
			\hline
			Modulation & $O(K^2)$ & \multicolumn{2}{c}{$O(K^2)$}\\
			\hline
			ML detection & $O((2^{\lfloor\log_{2}K! \rfloor}\cdot M^K)K^3)$ & $O((2^{\lfloor \log_{2}K! \rfloor}\cdot M)K^3)$ & $O((2^{\lfloor \log_{2}K! \rfloor}\cdot2 M)K^3)$\\
			\hline
		\end{tabular}
		\label{tb:DSTMcomplexity}
	\end{table}
	
	The results are summarized in Table \ref{tb:DSTMcomplexity}. We see that the complexity of the DRM-DSTM  scheme is not larger than that of the uncoded DRM for $M\geq2$ and $K\geq2$. Actually it is even lower in the mapping of information bits and ML detection processes.However, we need to remember that in coded DRM-DSTM we work with matrices, while for the uncoded DRM we work directly with $M$-PSK symbols. For example, the extraction of $M$-PSK symbols from decoded matrices require also resorces with a complexity depending  on $K$.
	
	\section{Simulation Result}\label{section4}

	In this section, we present Monte Carlo simulations for DRM and DRM-DSTM systems of larger sizes, where $K\geq2$. Acknowledging that very large sizes can greatly increase simulation time, we limit our simulations to the range $2\leq K\leq 4$. The setup of Monte Carlo simulations is the same as in section \ref{section:2.3}.
	
	\begin{figure}[htbp!]
		\centering
		\includegraphics[width=0.5\linewidth]{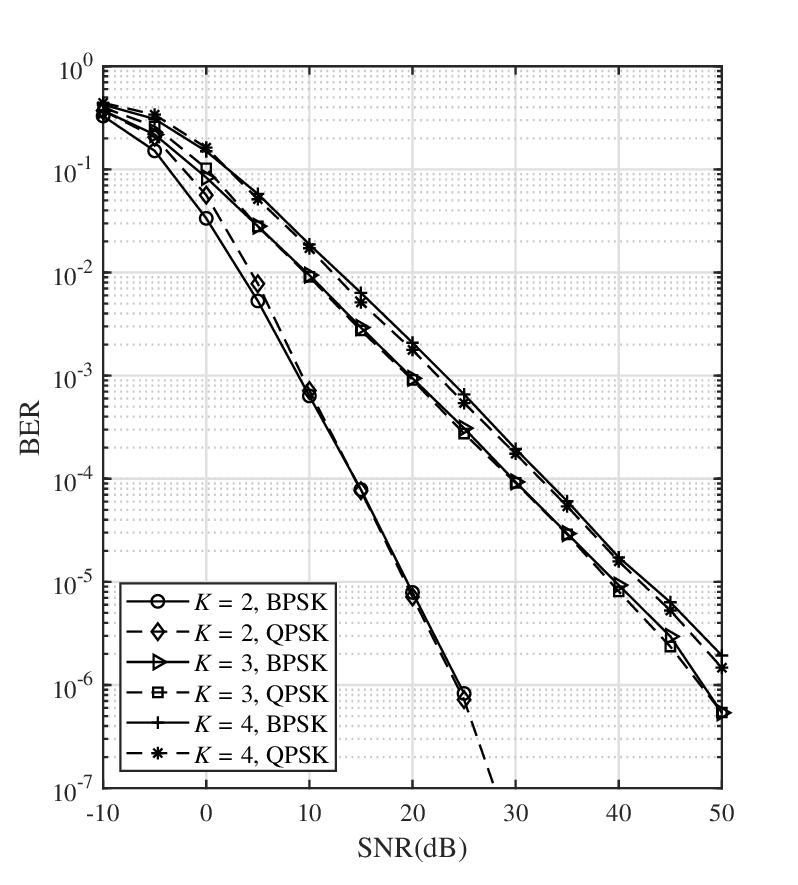}
		\caption{Performance of uncoded DRM for $2\leq K\leq 4$.}
		\label{Uncoded}
	\end{figure}

	In Fig. \ref{Uncoded} we present the performance of the original uncoded DRM scheme for $K$ values from 2 to 4.  At low SNR, it is evident that BPSK provides better error performance compared to QPSK. However, as the SNR increases, the performance of the two constellations gradually becomes closer. For instance, with $K=3$, BPSK outperforms QPSK until both schemes reach a BER of approximately $2.8\cdot10^{-2}$ at an $E_b/N_0$ of 5 dB. Beyond this point, the two curves almost overlap in the medium SNR range. For $K = 4$, the crossover point, where QPSK matches BPSK in error performance, occurs earlier, at a BER of around $10^{-1}$ when the SNR is 2.5 dB.
Furthermore, it is seen that the performance of the DRM scheme deteriorates significantly as $K$ increases. For instance, with BPSK at a BER of $10^{-5}$, the difference between $K=3$ and $K=4$ is around 2.5 dB, while the gap between $K=2$ and $K=3$ is approximately 20 dB. This trend highlights the negative impact of increasing $K$ on system performance.

	\begin{figure}[htbp!]
		\centering
		\includegraphics[width=0.5\linewidth]{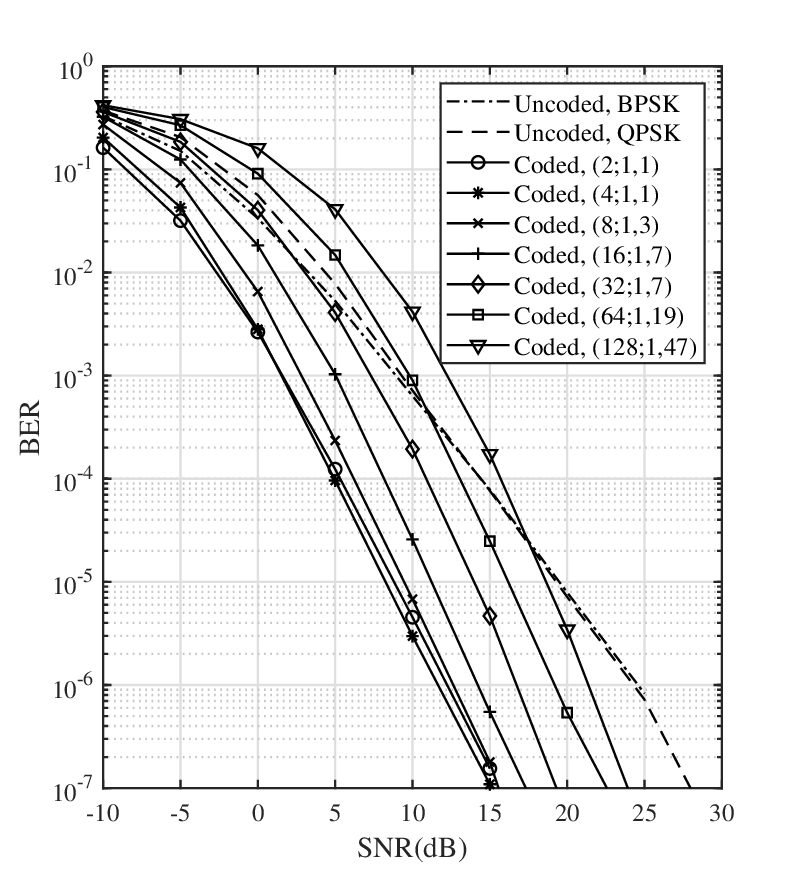}
		\caption{Performance of cyclic-coded scheme when $K=2$.}
		\label{cyclicK=2}
	\end{figure}
	
	\begin{figure}[htbp!]
		\centering
		\includegraphics[width=0.5\linewidth]{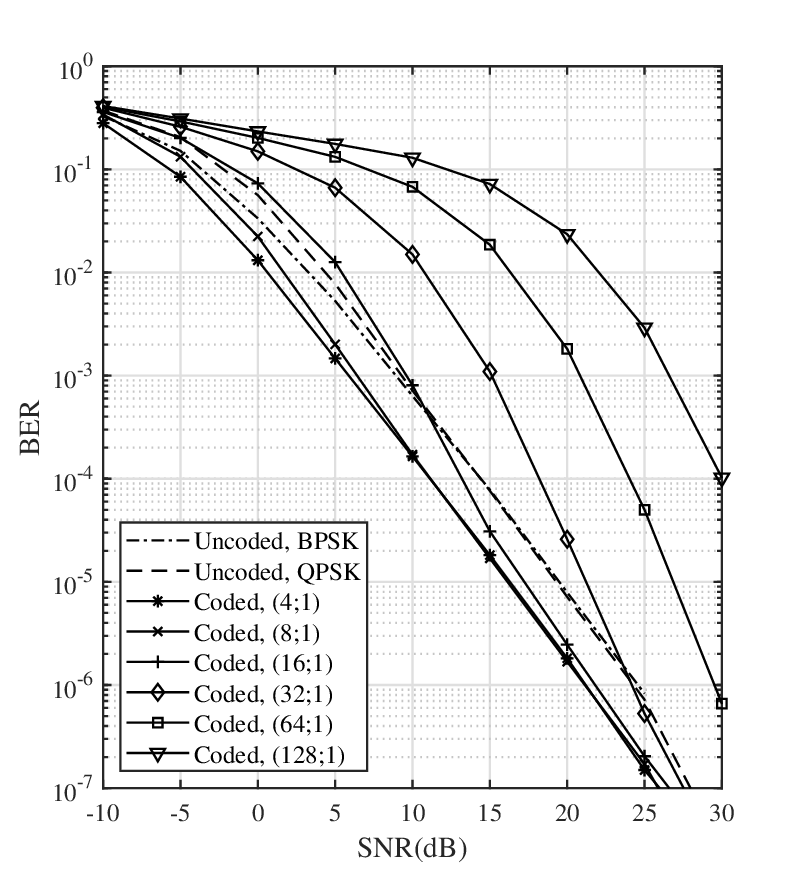}
		\caption{Performance of dicyclic-coded scheme when $K=2$.}
		\label{dicyclicK=2}
	\end{figure}
	
	In  Fig. \ref{cyclicK=2} we illustrate the performane of DRM-DSTM with cyclic-codes from Table \ref{tb:K=2}. As $M$ increases, BER performance worsens. Specifically, at a BER of $10^{-5}$, the $(8;1,3)$-coded scheme performs better than the $(16;1,7)$-coded scheme by about 2 dB and performs roughly 0.5 dB worse than the $(4;1,1)$-coded scheme. Moreover, when $M\geq16$, the SNR gap between successive cases becomes wider compared to the gap observed for $2\leq M\leq16$. For instance, the SNR difference between the cyclic codes $(64;1,19)$ and $(128;1,47)$ is about 4 dB at a BER of $10^{-5}$, which is larger than the 2 dB difference between the $(16;1,7)$-coded and $(8;1,3)$-coded schemes mentioned earlier. This suggests that as $M$ increases, the loss in error rate  performance increases.
Additionally, for $M\leq16$, the coded scheme consistently outperforms the DRM scheme across the entire SNR range. At a BER of $10^{-5}$, the $(2;1,1)$-coded scheme is roughly 10 dB better than the DRM scheme with BPSK, and the $(16;1,7)$-coded scheme surpasses uncoded DRM BPSK by around 8 dB. However, for $M\geq32$, the coded scheme initially performs worse than the uncoded DRM but eventually surpasses it as the SNR increases. For example, at a BER of $10^{-3}$, the original DRM with QPSK has a 2.5 dB advantage over the $(128;1,47)$-coded scheme. Nevertheless, at a BER of $10^{-5}$, the situation reverses, with uncoded DRM having a 0.5 dB disadvantage. 
	
	 In Fig. \ref{dicyclicK=2} we present error rate results for DRM-DSTM with the dicyclic codes from Table \ref{tb:K=2}.We see that the error rate performance deteriorates with increasing {M}, and the SNR gap at same BER becomes more pronounced compared to the cyclic-coded cases. For instance, the $(16;1)$-coded scheme outperforms the $(4;1)$-coded scheme by about 1 dB at a BER of $10^{-5}$. In contrast, the gap widens to around 4 dB between the $(32;1)$ and $(16;1)$ schemes, and to 6 dB between the $(64;1)$ and $(32;1)$ schemes. Additionally, only for $M=4$ and $8$, the coded schemes consistently surpass uncoded DRM. For $M=16$, the error rate is superior to uncoded DRM  with either constellation 
up to a BER of $4\cdot10^{-4}$ and  SNR of 11 dB. Beyond that point, the DRM- DSTM scheme performs worse compared to uncoded DRM. These observations underscore the importance of selecting an appropriate $M$ to prevent performance losses compared to uncoded DRM, even though the DRM-DSTM coded system is better in most scenarios.
	
	\begin{figure}[htbp!]
		\centering
		\includegraphics[width=0.5\linewidth]{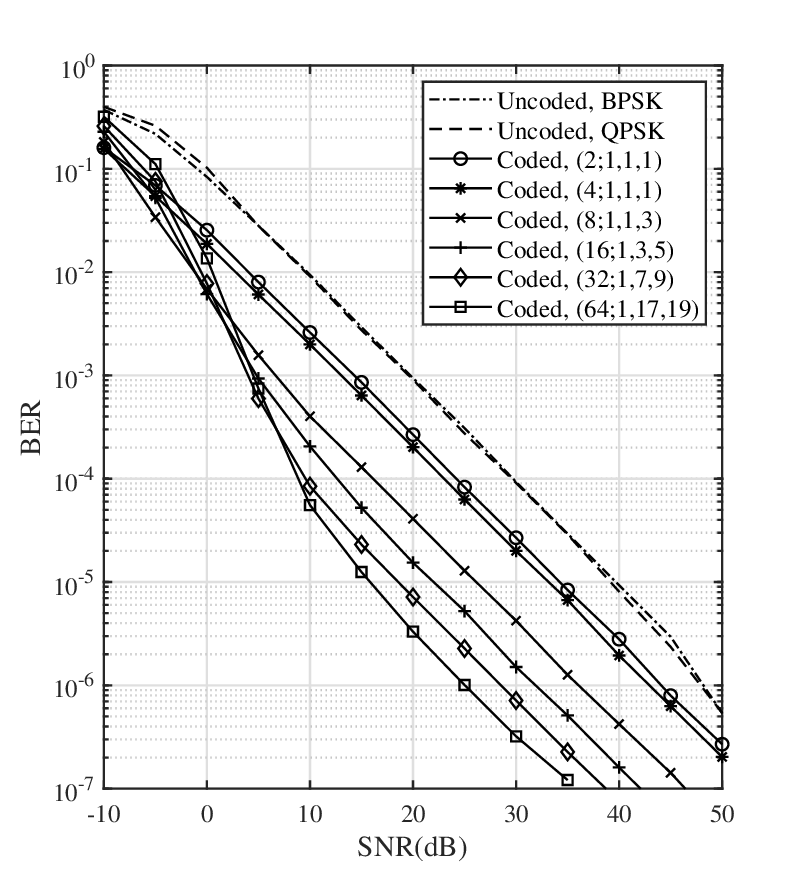}
		\caption{Performance of cyclic-coded scheme when $K=3$.}
		\label{cyclicK=3}
	\end{figure}
	
	Comparing the performance of cyclic and dicyclic codes with same $M$ indicates that cyclic codes significantly outperform the dicyclic ones. For example, the cyclic-coded scheme with $(32;1,7)$ achieves a BER of $10^{-5}$ at an $E_b/N_0$ of around 14 dB, while the corresponding dicyclic-coded scheme with $(32;1)$ requires approximately 21 dB to reach the same BER. This performance gap widens as $M$ increases. Specifically, at a BER of $10^{-4}$, the $(128;1,47)$ cyclic system needs only 15.5 dB SNR, whereas the dicyclic system requires 30 dB. This emphasizes how structural differences between cyclic and dicyclic codes can significantly affect performance, with the gap in BER performance becoming more evident as $M$ grows. 
	
In Fig. \ref{cyclicK=3} we present the BER performance of the cyclic codes listed in Table \ref{tb:K=3}. In contrast to the trends we found for $K=2$, this figure reveals a performance improvement with increasing $M$ for all values considered in this work. For instance, at a BER of $10^{-5}$, the $(64;1,17,19)$-coded scheme shows the best performance, surpassing the second-best $(32;1,7,9)$ scheme by roughly 1 dB and outperforming the worst-performing $(2;1,1,1)$ scheme by about 20 dB. Additionally, the error performance of all the considered cyclic codes  significantly exceeds that of  uncoded DRM  with BPSK and QPSK for  same $K$. At a BER of $10^{-5}$, all the coded schemes in Fig. \ref{cyclicK=3} perform better than  DRM  by roughly 5 to 25 dB. This substantial gap underscores the superiority of the coded system when $K=3$. As $M$ increases, the error rate performance consistently improves. This advantage becomes particularly clear at an SNR larger than 10 dB..

	\begin{figure}[htbp!]
		\centering
		\includegraphics[width=0.5\linewidth]{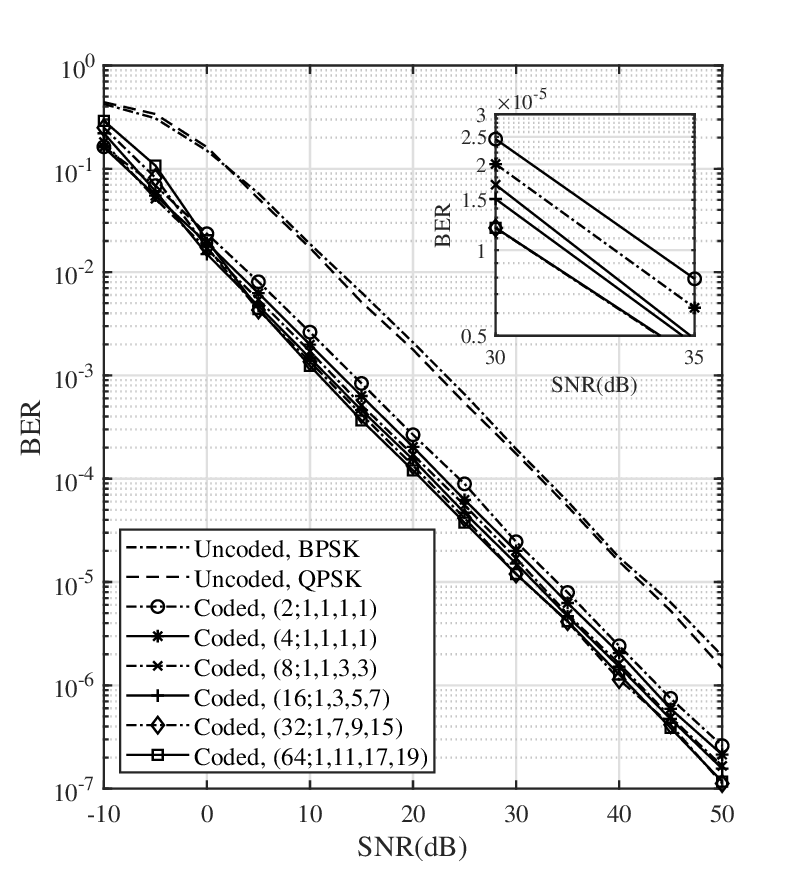}
		\caption{Performance of cyclic-coded scheme when $K=4$.}
		\label{cyclicK=4}
	\end{figure}
	
\begin{figure}[htbp!]
		\centering
		\includegraphics[width=0.5\linewidth]{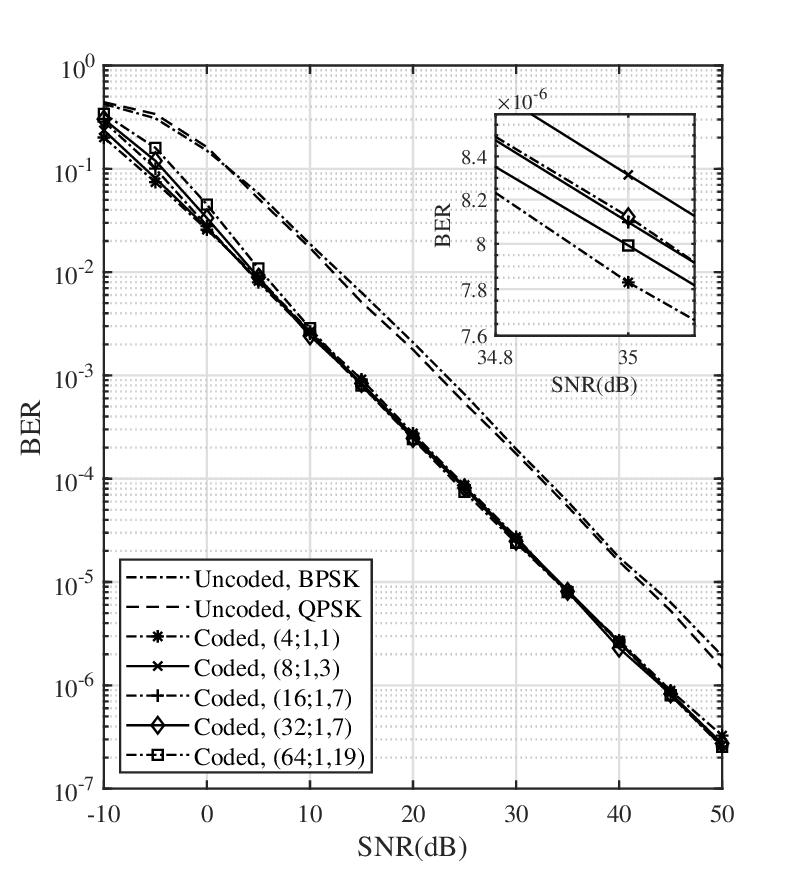}
		\caption{Performance of dicyclic-coded scheme when $K=4$.}
		\label{dicyclicK=4}
	\end{figure}

	In Fig. \ref{cyclicK=4} and \ref{dicyclicK=4} we present the BER performance of the cyclic and dicyclic group codes from Table \ref{tb:K=4}. Both figures compare these coded schemes with the uncoded DRM systems using BPSK and QPSK. Across the entire SNR range, every coded system outperforms the uncoded schemes for various $M$. For instance, at a BER of $10^{-5}$, the $(4;1,1,1,1)$ cyclic-coded scheme shows about an 8 dB improvement over the uncoded BPSK system. This gap widens to 12 dB between the $(64;1,11,17,19)$ cyclic-coded scheme and the uncoded BPSK system. Similarly, when the BER is $10^{-5}$, all the dicyclic-coded schemes surpass the DRM scheme with QPSK by approximately 10 dB. These results indicate the advantages of DRM-DSTM systems over uncoded DRM for $K=4$. The increasing gap with larger $M$ further suggests the enhanced error-correction capabilities as $M$ grows. Furthermore, for the cyclic codes in Fig. \ref{cyclicK=4}, similar trends to those observed for $K=3$ is seen. When $M$ increases, BER performance improves, though the performance gap between different scenarios is smaller than for $K=3$. At a BER of $10^{-5}$, the $(32;1,7,9,15)$ scheme outperforms the $(16;1,3,5,7)$  scheme by about 1 dB, and the $(16;1,3,5,7)$ scheme surpasses the $(8;1,1,3,3)$ scheme by approximately 0.5 dB. Although these gaps are smaller than the differences observed for $K=3$, they still indicate the positive impact of increasing $M$ on system performance.
	
	In Fig. \ref{dicyclicK=4} we present results for dicyclic coded DRM-DSTM with $K=4$.  We notice even smaller gaps between different codes, with the curves nearly overlapping at high SNR. For example, at a BER of $8\cdot10^{-6}$, the $(4;1,1)$ dicyclic-coded scheme outperforms the $(64;1,19)$ scheme by just 0.1 dB and exceeds the $(16;1,7)$ scheme by only 0.15 dB. In addition, the performance gap between cyclic and dicyclic codes decreases as compared to $K=2$, especially at high SNR. For instance, at a BER of $10^{-5}$, the $(64;1,11,17,19)$ cyclic scheme in Fig. \ref{cyclicK=4} surpasses the $(64;1,19)$ dicyclic scheme in Fig. \ref{dicyclicK=4} by only 3 dB, indicating a significant reduction from the 10 dB difference observed for $K=2$. This demonstrates that cyclic codes still offer more robust error performance for $K=4$, but the advantage diminishes compared to dicyclic codes.
	
	These results show that while increasing code lengths continue to enhance performance, the impact becomes less pronounced as $K$ increases. The nearly overlapping curves of the dicyclic codes indicate that beyond a certain point, the performance difference becomes minimal, especially at high SNR. The reduced performance gap between cyclic and dicyclic codes for $K=4$ compared to $K=2$ further suggests that the advantage of cyclic codes diminishes as $K$ grows, leading to a convergence in the effectiveness of both coding schemes.
	
	\section{Conclusion}
	This paper introduces DRM-DSTM which is a DSTM coded  DRM technique  to bypass the need for channel estimation in RIS communication systems. We consider an RIS-based DRM scheme, featuring $K$ legitimate reflecting patterns and $K$ symbol slots in each block. We introduce the fundamentals of DSTM and explore methods for integrating this coding technique  with DRM. The performance of both uncoded and coded systems is initially presented for $K=2$, demonstrating the feasibility and superiority of our coded technique.
We extend the original DRM and coded scheme to higher values of  $K$ and  present important general aspects for the design of coded DRM-DSTM schems. Extensive BER simulation results over quasi-static Rayleigh fading channels  indicate the effectiveness of this technique for various dimensions and its superiority over uncoded DRM in a majority of scenarios, when selecting proper design parameters. Our work shows that although the performance improves with increasing $K$, beyond $K=4$ the returns diminish.
This paper focuses primarily on cyclic and dicyclic group codes for DRM-DSTM, however it is important to note that other types of codes, such as those based on orthogonal designs, can be also accomodated in our framework constituting a worthwile subject for future investigations.

	\appendices
	\section{Derivation of the Signal Model}\label{apdx1}
	We first prove $\mathbf{V}[t]=\tilde{\mathbf{Z}}[t]\tilde{\mathbf{S}}[t]$, where $\tilde{\mathbf{Z}}[t]$ is a permutation matrix and $\tilde{\mathbf{S}}[t]$ is a diagonal matrix. It is known that multiplying diagonal matrices results in another diagonal matrix, and this also holds true for permutation matrices. Furthermore, any permutation matrix $\mathbf{P}$ satisfies $\mathbf{P}\mathbf{P}^{\rm T}=\mathbf{P}^{\rm T}\mathbf{P}=\mathbf{I}$. Therefore, the product of a diagonal matrix $\mathbf{D}$ and a permutation matrix $\mathbf{P}$ of the same size can be expressed as $\mathbf{D}\mathbf{P}=\mathbf{P}\mathbf{P}^{\rm T}\mathbf{D}\mathbf{P}=\mathbf{P}\mathbf{D}'$, where $\mathbf{D}'=\mathbf{P}^{\rm T}\mathbf{D}\mathbf{P}$ is also diagonal, and its entries are permuted by $\mathbf{P}$. Using this property, we arrive at $\mathbf{P}_1\mathbf{D}_1\mathbf{P}_2\mathbf{D}_2=\mathbf{P}_1\mathbf{P}_2\mathbf{D}_1'\mathbf{D}_2=\tilde{\mathbf{P}}_2\tilde{\mathbf{D}}_2$, where $\tilde{\mathbf{P}}_2=\mathbf{P}_1\mathbf{P}_2$ is a permutation matrix, and $\tilde{\mathbf{D}}_2=\mathbf{D}_1'\mathbf{D}_2=\mathbf{P}_2^{\rm T}\mathbf{D}_1\mathbf{P}_2\mathbf{D}_2$ is a diagonal matrix. Then, following (\ref{eq:diffencode}), we have 
	\begin{align}
		\mathbf{V}[1] =&\mathbf{V}[0]\mathbf{Z}[1]\mathbf{S}[1]=\mathbf{Z}[1]\mathbf{S}[1], \notag\\ \mathbf{V}[2]=&\mathbf{Z}[1]\mathbf{S}[1]\mathbf{Z}[2]\mathbf{S}[2]=(\mathbf{Z}[1]\mathbf{Z}[2])(\mathbf{Z}[2]^T\mathbf{S}[1]\mathbf{Z}[2]\mathbf{S}[2])=\tilde{\mathbf{Z}}[2]\tilde{\mathbf{S}}[2], \notag\\ \mathbf{V}[3]=&\tilde{\mathbf{Z}}[2]\tilde{\mathbf{S}}[2]\mathbf{Z}[3]\mathbf{S}[3]=(\tilde{\mathbf{Z}}[2]\mathbf{Z}[3])(\mathbf{Z}[3]^T\tilde{\mathbf{S}}[2]\mathbf{Z}[3]\mathbf{S}[3])=\tilde{\mathbf{Z}}[3]\tilde{\mathbf{S}}[3], \notag\\
		\cdots&, \notag\\
		\mathbf{V}[t]=&\tilde{\mathbf{Z}}[t-1]\tilde{\mathbf{S}}[t-1]\mathbf{Z}[t]\mathbf{S}[t]=(\tilde{\mathbf{Z}}[t-1]\mathbf{Z}[t])(\mathbf{Z}[t]^T\tilde{\mathbf{S}}[t-1]\mathbf{Z}[t]\mathbf{S}[t])=\tilde{\mathbf{Z}}[t]\tilde{\mathbf{S}}[t],
	\end{align}
	where 
	\begin{align}
		\tilde{\mathbf{Z}}[1]&=\mathbf{Z}[1], \notag\\
		\tilde{\mathbf{Z}}[2]&=\mathbf{Z}[1]\mathbf{Z}[2], \notag\\
		\tilde{\mathbf{Z}}[3]&=\mathbf{Z}[1]\mathbf{Z}[2]\mathbf{Z}[3], \notag\\
		\cdots&, \notag\\
		\tilde{\mathbf{Z}}[t]&=\mathbf{Z}[1]\mathbf{Z}[2]\cdots\mathbf{Z}[t],
	\end{align}
	and
	\begin{align}
		\tilde{\mathbf{S}}[1]&=\mathbf{S}[1],\notag\\
		\tilde{\mathbf{S}}[2]&=\mathbf{Z}[2]^T\mathbf{S}[1]\mathbf{Z}[2]\mathbf{S}[2],\notag\\
		\tilde{\mathbf{S}}[3]&=\mathbf{Z}[3]^T\tilde{\mathbf{S}}[2]\mathbf{Z}[3]\mathbf{S}[3]=\mathbf{Z}[3]^T\mathbf{Z}[2]^T\mathbf{S}[1]\mathbf{Z}[2]\mathbf{S}[2]\mathbf{Z}[3]\mathbf{S}[3],\notag\\
		\cdots&,\notag\\
		\tilde{\mathbf{S}}[t]&=\mathbf{Z}[t]^T\cdots\mathbf{Z}[2]^T\mathbf{S}[1]\mathbf{Z}[2]\mathbf{S}[2]\cdots\mathbf{Z}[t]\mathbf{S}[t].
	\end{align}
	Thus, both $\tilde{\mathbf{Z}}[t]$ and $\tilde{\mathbf{S}}[t]$ remain permutation and diagonal matrices, respectively.
	
	\section{Derivation of the Differential Detection Method}\label{apdx2}
	Following (\ref{eq:MLdetect}), we present the detailed derivation of the ML detection technique for our system.
	\begin{align}
		\hat {\mathbf {X}}[t]&=\arg \min \limits_{\mathbf {X}[t]\in \mathcal {X}}\lVert\mathbf {Y}[t]-\mathbf {Y}[t-1]\mathbf {X}[t]\rVert_{F}^{2} \notag\\
		&=\arg \min \limits_{\mathbf {X}[t]\in \mathcal {X}}{\rm {Tr}}\{(\mathbf{Y}[t]-\mathbf{Y}[t-1]\mathbf{X}[t])^{H}(\mathbf{Y}[t]-\mathbf{Y}[t-1]\mathbf{X}[t])\} \notag\\
		&=\arg \min \limits_{\mathbf {X}[t]\in \mathcal {X}}{\rm {Tr}}\{(\mathbf{Y}^{H}[t]-\mathbf{X}^{H}[t]\mathbf{Y}^{H}[t-1])(\mathbf {Y}[t]-\mathbf {Y}[t-1]\mathbf{X}[t])\}\notag\\
		&=\arg \min \limits_{\mathbf {X}[t]\in \mathcal {X}}\Big\{{\rm {Tr}} \{\mathbf{Y}^{H}[t]\mathbf{Y}[t]+\mathbf{X}^{H}[t]\mathbf{Y}^{H}[t-1]\mathbf{Y}[t-1]\mathbf{X}[t] \notag\\
		&-\mathbf{Y}^{H}[t]\mathbf{Y}[t-1]\mathbf{X}[t]-\mathbf{X}^{H}[t]\mathbf{Y}^{H}[t-1]\mathbf{Y}[t]\}\Big\}	\notag\\
		&=\arg \min \limits_{\mathbf {X}[t]\in \mathcal {X}}\Big\{\lVert\mathbf{Y}[t-1]\mathbf{X}[t]\rVert^2_F-{\rm{Tr}}\{ {(\mathbf{Y}^{H}[t]\mathbf{Y}[t-1]\mathbf{X}[t])+(\mathbf{Y}^{H}[t]\mathbf{Y}[t-1]\mathbf{X}[t])^{H}}\}\Big\}\notag\\
		&=\arg \min \limits_{\mathbf {X}[t]\in \mathcal {X}} \Big\{\lVert\mathbf{Y}[t-1]\mathbf{X}[t]\rVert^2_F-{\rm{Tr}}\{\mathbf{Y}^{H}[t]\mathbf{Y}[t-1]\mathbf{X}[t]\}-{\rm{Tr}}\{(\mathbf{Y}^{H}[t]\mathbf{Y}[t-1]\mathbf{X}[t])^{H}\}\Big\}\notag\\
		&=\arg \min \limits_{\mathbf {X}[t]\in \mathcal {X}}\Big\{\lVert\mathbf{Y}[t-1]\mathbf{X}[t]\rVert^2_F-2\Re \{ {\rm{Tr}} \{\mathbf{Y}^{H}[t]\mathbf{Y}[t-1]\mathbf{X}[t]\}\}\Big\} \label{eq:penultimate}\\
		&=\arg \max \limits_{\mathbf {X}[t]\in \mathcal {X}} \Re \left \{{\rm {Tr}}(\mathbf {Y}^{H}[t]\mathbf {Y}[t-1]\mathbf {X}[t])\right \}, \label{eq:laststep}
	\end{align}
	where $\mathcal{X}$ is the set of all legitimate $\mathbf{X}_t$ and $|\mathcal{X}|=2^r$. The relation in  (\ref{eq:penultimate}) can be proved by the following steps. The trace of any $n\times n$ square matrix $\mathbf{A}$ is defined as ${\rm {Tr}}(\mathbf{A})=\sum_{i=1}^{n}a_{ii}=a_{11}+a_{22}+\cdots+a_{nn}$ and ${\rm {Tr}}(\mathbf{A}^H)={\rm {Tr}}(\mathbf{A})^*$. Therefore, ${\rm {Tr}}(\mathbf{A})+{\rm {Tr}}(\mathbf{A}^{H})={\rm {Tr}}(\mathbf{A})+{\rm {Tr}}(\mathbf{A})^*=2\Re\{{\rm {Tr}}(\mathbf{A})\}$. Since $ {\mathbf{Y}^{H}[t]\mathbf{Y}[t-1]\mathbf{X}[t]}\in\mathbb{C}^{K\times K}$ is a square matrix, then ${\rm{Tr}} \{\mathbf{Y}^{H}[t]\mathbf{Y}[t-1]\mathbf{X}[t]\}+{\rm{Tr}}\{(\mathbf{Y}^{H}[t]\mathbf{Y}[t-1]\mathbf{X}[t])^{H}\}=2\Re \{ {\rm{Tr}} \{\mathbf{Y}^{H}[t]\mathbf{Y}[t-1]\mathbf{X}[t] \}\}$. 
	
	Next, to obtain (\ref{eq:laststep}), the term $\lVert\mathbf{Y}[t-1]\mathbf{X}[t]\rVert^2_F$ can be expressed as
	\begin{align}
		\lVert\mathbf{Y}[t-1]\mathbf{X}[t]\rVert^2_F=&{\rm Tr}\{\mathbf{Y}[t-1]\mathbf{X}[t]\mathbf{X}^{H}[t]\mathbf{Y}^{H}[t-1]\} \label{eq:trYY1}\\
		=&{\rm Tr}\{\mathbf{Y}[t-1]\mathbf{Y}^{H}[t-1]\}, \label{eq:trYY2}
	\end{align}
	where $\mathbf{X}[t]=\mathbf{Z}[t]\mathbf{S}[t]$ is defined in (\ref{eq:InforCarry}). The permutation matrix $\mathbf{Z}[t]$ is orthogonal, resulting in $\mathbf{Z}^H[t]=\mathbf{Z}^{-1}[t]$ and $\mathbf{Z}^H[t]\mathbf{Z}[t]=\mathbf{I}$. The diagonal matrix $\mathbf{S}[t]$ with diagonal elements being $M$-PSK symbols is also unitary, having $\mathbf{S}^{H}[t]\mathbf{S}[t]=\mathbf{I}$. Then, $\mathbf{X}[t]\mathbf{X}^{H}[t]=\mathbf{Z}[t]\mathbf{S}[t]\mathbf{S}^H[t]\mathbf{Z}^H[t]=\mathbf{I}$ resulting in (\ref{eq:trYY2}). Therefore, $\lVert\mathbf{Y}[t-1]\mathbf{X}[t]\rVert^2_F$ can be ignored in the decision statistc because it does not depend on  $\mathbf{X}[t]\in\mathcal{X}$ when $\mathbf{Y}[t-1]$ is given.
	Finally, $\hat{\mathbf{X}}[t]$ can be decoded according to the mapping rule of $\mathbf{X}[t]$.

	\bibliographystyle{IEEEtran}
	\bibliography{reference}

\end{document}